%% file: paper4_LMCS.tex
\documentclass{lmcs}
\pdfoutput=1

\usepackage{lastpage}
\lmcsdoi{15}{1}{32}
\lmcsheading{}{\pageref{LastPage}}{}{}%
{Oct.~30,~2017}{Mar.~29,~2019}{}

\keywords{Groupoids,
Networks,
Quantum,
Semantics,
Key distribution}

\input{header_LMCS}
\tikzset{blob/.style={draw, circle, fill=white, inner sep=1pt, minimum width=10pt, font=\scriptsize, node on layer=front, thick}}

\tikzset{classical/.style={draw=none, fill=yellow!80!red, fill opacity=0.5}}

\begin{document}

\title{A classical groupoid model for quantum networks}
\email{david.reutter@cs.ox.ac.uk, j.o.vicary@bham.ac.uk}

\author[D.J. Reutter]{David J. Reutter\rsuper{a}}
\address{\lsuper{a}Department of Computer Science, University of Oxford}
\author[J. Vicary]{Jamie Vicary\rsuper{{a,b}}}
\address{\lsuper{b}School of Computer Science, University of Birmingham}


\begin{abstract}
We give a mathematical analysis of a new type of classical computer network architecture, intended as a model of a new technology that has recently been proposed in industry. Our approach is based on groubits, generalizations of classical bits based on groupoids. This network architecture allows the direct execution of a number of protocols that are usually associated with quantum networks, including teleportation, dense coding and secure key distribution.

\end{abstract}

\maketitle

\section{Introduction}

\label{sec:overview}

\noindent
Borrill and Karp have recently introduced the notion of \emph{timeless network}~\cite{Borrill:2016}, a new paradigm for distributed communication currently under commercial development by Earth Computing\footnote{See \href{http://www.earthcomputing.io}{http://www.earthcomputing.io}.}. Inspired by their proposal, we introduce a new 
network architecture based on \textit{groubits}---group-theoretical generalizations of classical bits, with similar behaviour to qubits in quantum information---and go on to show that groubits can be manipulated to achieve a wide range of surprising informatic tasks. We give a categorical syntax and semantics for groubits, and develop a graphical calculus to prove correctness of groubit protocols.

\paragraph{Groubits.} A groubit is a computational device storing two ordinary bits $(A_L, A_I)$, a \emph{logical bit} $A_L$ and an \emph{internal bit} $A_I$, and supporting the primitive operations \Init, \Swap, \Read, \Write and \Tick. Some of these operations in turn make use of the procedure  $\Rand$, a function with no arguments which returns either  0 or 1 nondeterministically. We describe these procedures as follows. The \Init operation takes no arguments, and creates a new groubit in the following state:
\begin{itemize}
  \item $\Init = (\Rand, 0)$
\end{itemize}
Here and throughout, we intend that the $\Rand$ function is executed freshly each time. The \Swap operation acts on a groubit, exchanging the logical and internal bits:
\begin{itemize}
  \item $\Swap(A_L, A_I) = (A_I, A_L)$
\end{itemize}
Conventional single bits $[B]$ can be stored in groubits, using the following read and write procedures:
\begin{itemize}
  \item $\Read(A_L, A_I) = [A_L]$
  \item $\Write[B] = (B, \Rand)$
\end{itemize}
The \Read operation destroys a groubit and creates a conventional bit,  while the \Write operation destroys a conventional bit and creates a new groubit. By `destroy', we mean that the corresponding structure is no longer available for interaction; of course, in a real-world implementation, it may not be physically destroyed, but rather have its informational content somehow rendered inaccessible. Pairs of groubits can also be connected by a \textit{link}, enabling the \Tick  operation, where $A$ and $B$ label the two connected nodes, and $\oplus$ is addition modulo~2:
{\def\,{\hspace{0.6pt}}
\begin{itemize}
        \item $\Tick((A_L{,\,}A_I){,\,}(B_L{,\,}B_I) \big)
        {\,=\,}((A_L{,\,} A_I \soplus B_L){,\,}(B_L{,\,} B_I \soplus A_L))$
\end{itemize}}

\noindent
Intuitively, for each node in the pair, we flip the internal bit just when the other node has logical bit equal to 1. Nodes can belong to multiple links, forming a graph topology.

\paragraph{Assumptions.}
\label{sec:assumptions}
We make some assumptions about these groubit operations.
\begin{itemize}
        \item \textbf{Atomicity.} The operations \Init, \Swap, \Read, \Write and \Tick are atomic; that is, they either succeed or have no effect at all, with the parties involved being aware which of these two possibilities has occurred.
         \item \textbf{Security.} The state of a node cannot be accessed, except via \Read.
\end{itemize}
We emphasize that claims we make about the functionality of groubit networks---in particular, security properties---rest on the validity of these assumptions.\footnote{For quantum protocols such as quantum key distribution, security is derivable from the laws of physics~\cite{Vazirani_2014}; this is not the case here.} We suggest that these assumptions are within the realm of technological plausibility; for example, separation kernels~\cite{Zhao:2015} are a well-developed technology for guaranteeing strong security properties of private memory states within embedded devices. Our focus here is on the logical properties of these devices, rather than on questions of implementation, so we do not discuss these aspects further. Note  however that we do not assume that devices cannot fail; to satisfy the assumptions, it would be valid for a device to self-destruct if tampering was detected.

\subsection{Significance}

\noindent
We claim that groubits have exotic properties making them interesting to study. In particular, they allow timeout-free atomic message routing (the origin of the term `timeless network'), and they have the ability to replicate a variety of quantum protocols.

\paragraph{Message routing.} Linear chains of groubits allow message routing between nodes with guaranteed message atomicity, and without timeouts (see Section~\ref{sec:statetransfer}). We understand that developing this idea is the primary commercial interest of Earth Computing, with a focus on high-resilience network architectures for data centres; this is potentially significant, since the timeout properties of the standard TCP transport protocol~\cite{Iren_1999} can cause reliability issues in a data centre environment~\cite{Adesanmi_2015, Borrill:2016}.

\paragraph{Quantum behaviour.}
A range of quantum protocols---entanglement creation, teleportation, dense coding, and secure key distribution---can be implemented on a groubit network, almost without modification.

If groubit networks can be implemented at scale in the real world, this may prove technologically significant, given the possibility that quantum computers may within decades be able to break in polynomial time the RSA public-key encryption scheme which is currently technologically dominant~\cite{Bernstein:2009}. Should this possibility be realized, it has been suggested that quantum key distribution could be used as an alternative technology to enable long-range information--theoretically secure communication~\cite{Alleaume:2007}; we suggest that key distribution running on a large-scale groubit network may be an alternative worth investigating.

Some points must be made completely clear. Information--theoretically secure key distribution is known to be impossible in a classical computation setting. Our claim that it can be implemented using networks of groubits rests on the atomicity and security assumptions given in Section~\ref{sec:assumptions}, and will hold for any real-world implementation only to some approximation. Also, we do not claim that \emph{all} quantum protocols or algorithms can be implemented on groubits; in particular, we expect no analogue of `quantum speedup', and give no classical model for important procedures such as the Grover or Shor algorithms~\cite{Nielsen:2009}.

Nonetheless, for those quantum protocols that we claim can operate on a groubit network, we mean this in a strong sense. In~\autoref{sec:quantization}, we present a quantization functor which gives a structure-preserving mapping from our setting into quantum theory, sending groubit protocols to quantum protocols, and sending a groubit to a Hadamard matrix~\cite{Reutter:2016,Vicary:2012}. In other words, groubits yield a local hidden variable model for the part of quantum theory in the image of this quantization functor.


\subsection{Overview}

The structure of this article is as follows. In Section~\ref{sec:foundations},  we give the definition of a groubit in terms of groupoids with extra structure. We define the 2\-category \GAf of finite groupoids, free profunctors and spans, and in our central technical result, show that groubits correspond precisely to {biunitary connections} in \GAf\footnote{See Section~\ref{sec:relatedwork} for background on biunitaries.}. We also give a 2\-dimensional graphical programming language for groubits, and give a thorough development of its syntax and semantics. In Section~\ref{sec:protocols} we give programs for state transfer, entanglement creation, teleportation and dense coding on networks of groubits, and verify these protocols using the rules of our abstract 2\-dimensional syntax. We comment on the potential applicability of these protocols for message transfer and key distribution within networks of groubits. Further technical details on \GAf and its quantization functor are given in~\autoref{sec:bicategories}.
 
\subsection{Related work}
\label{sec:relatedwork}

\firstparagraph{Timeless networks.}
The concept of timeless networking and the possibility of timeout-free atomic message routing is due to Borrill and Karp~\cite{Borrill:2016}, who also described the quantum properties of the technology. Our treatment here is inspired in part by their ideas, but does not follow the technical details of their approach.


\paragraph{Spekkens' toy model.} A toy model for quantum phenomena has been developed by Spekkens and others~\cite{Spekkens:2007, Coecke_2011, Backens:2015, Disilvestro:2016, Pusey:2012} based on the \textit{knowledge balance} principle, in which quantum-like effects arise by restricting an observer's ability to gain information about the state of a classical system. This principle  can be seen as playing a role here, since groubits exhibit precisely such a balance between observable and unobservable states. Work on the toy model includes classical versions of several quantum procedures, including teleportation and dense coding  which we also analyze here; furthermore, the low-level combinatorics are strongly similar in places (compare for example \cite[Section I]{Spekkens:2007} with \autoref{fig:densecoding} here.)

Our work goes beyond these results in a number of ways, including: identification of biunitary structures in \GAf as a mathematical foundation; classification of these structures in terms of groubits; applications to timeless networks, key distribution and state transfer; the 2\-dimensional high-level language for designing and verifying groubit programs; and the identification of a functorial mapping from our calculus to quantum theory. Also, we have a fundamentally different perspective: while the work cited above studies the toy model as a `foil theory'---an exercise in quantum foundations which sheds light on the distinction between quantum and classical reality---our perspective is technological, focussed on writing and verifying programs for these hypothetical devices, which may be implementable and practically useful in the real world.

\paragraph{Groupoidification.} Our work is close in spirit to the groupoidification programme developed by Baez, Morton and others~\cite{Baez:2001, Baez:2010, Morton:2006, Bergeron_1997} from the combinatorial species of Joyal~\cite{Joyal_1981}; as here, they develop a 2\-categorical groupoid-based model for quantum-like phenomena, equipped with a functorial mapping into traditional quantum theory. Yet there is a surprising disconnect: their work is based on groupoids, spans, and spans of spans, while ours is based on groupoids, free profunctors and spans. This technical distinction seems mild, yet is essential for our results, and we are not aware of a direct relationship between the settings.

\paragraph{Classical key distribution.} Maurer~\cite{Maurer:1993} has suggested classical procedures for secure key distribution based on noisy communication channels. In his words, he drops the ``apparently innocent assumption that, except for the secret key, the enemy has access to precisely the same information as the legitimate receiver''. This is fundamentally different to our model, in which---just as in quantum key distribution---the ``enemy'' has access to the entire apparatus.

\paragraph{Biunitaries.} Our main proof technique is the technology of \textit{biunitaries} (see Section~\ref{sec:biunitaries}.) Introduced by Ocneanu~\cite{Ocneanu:1989} in 1989 and since developed by Jones, Morrison and others~\cite{Jones:1999,Jones:2013,Morrison:2014}, they are  a central tool in the classification of subfactors, a major research effort in pure mathematics. Biunitaries belong to the theory of \textit{planar algebras}, which studies the linear representation theory of algebraic structures in the plane. The 2\-dimensional syntax we use in this paper derives heavily from the work of this community. These planar algebra techniques have been used by the present authors and others~\cite{Vicary:2012, Reutter:2016, Reutter:2017a, Jaffe:2016a} to give a high-level language for quantum computation.

\paragraph{Unpublished work.} Related ideas have been described by Bar and the second author in an unpublished note~\cite{Bar:2014}.

\subsection{Acknowledgements}

\noindent
We are grateful to Paul Borrill and Alan Karp for conversations about timeless networking, Steve Vickers for many helpful comments on an earlier version of this paper, and to Krzysztof Bar for substantial discussions about groupoid semantics and a classical model for key distribution.

\section{Foundations}
\label{sec:foundations}

\subsection{Groudits and dits}
\label{sec:grouditsanddits}

\firstparagraph{Groudits.}
We begin with the definition of a groudit. Here and throughout, given a groupoid \cat G with a chosen object $a$, we write $\Aut_\cat G(a)$ for the set of morphisms in \cat G of type $a \to a$.

\begin{definition}
A \textit{groudit} $\mathcal G$ is a skeletal groupoid of the form \mbox{$\cat G = \coprod_{a} G_a$}, where $G_a$ are finite groups, equipped for each \mbox{$a \in \Ob(\cat G)$} with bijections $\epsilon_a, \tau_a : \Aut_\G(a) \to \Ob(\cat G) $.
\end{definition}

\noindent
Thinking about the consequences of this definition, we see that the underlying groupoid of a groudit is a disjoint union of $d$ finite groups for some $d \in \N$, each with $d$ elements. Note that the bijection data is not required to satisfy any properties, so groudits are easy to construct.

Our main result of this section is Theorem~\ref{thm:biunitaryclassification}, where we show that groudits classify biunitary structures in a 2\-category of finite groupoids, free profunctors and invariant spans; in particular, this theorem shows how for every groudit we can obtain analogues of the \Swap and \Tick maps, which we preview here as follows, using a notation $(a,b)$ where $a \in \Ob(\cat G)$ and $b:a \to a$:
\begin{align}
\Swap_\mathcal G (a,b) &:=\left( \epsilon_a(b), \tau ^{-1} _{\epsilon _{a} (b)}(a)\right)
\\
\Tick _\mathcal G((a,b),(c,d)) &:= \left(\vphantom{\frac{a}{b}} (a,b \left(\epsilon_a^{-1}(c) \right)^{-1}), \left(c, \tau_c^{-1}(a)d \right) \right)
\end{align}
\noindent
Much later, in Section~\ref{sec:quantization}, we show that groudits yield Hadamard matrices, quantum combinatorial structures of deep importance in quantum information; in this sense, groudits are the classical combinatorial analogues of Hadamard matrices.

\begin{figure}[b]
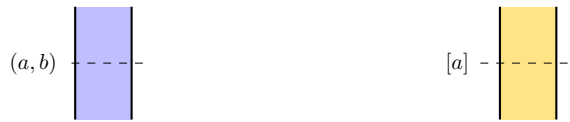

\figuretopsuck
\begin{calign}
\nonumber
\begin{tz}[scale=0.75,yscale=-1]
\draw [surface] (0,0) rectangle +(1,2);
\draw [edge] (0,0) to (0,2);
\draw [edge] (1,0) to +(0,2);
\def\leftmost{-1.3}
\foreach \y/\l/\t in
{
1/{(a,b)}/{}}
{
\node [programlabel] (1) at (\leftmost, \y) {$\l$};
\draw [hor] (1.east -| 1.2,0) to (1.east);
\node [programlabel] at (\leftmost, \y-0.5) {\t};
}
\end{tz}
&
\begin{tz}[scale=0.75,yscale=-1]
\draw [classical] (0,0) rectangle +(1,2);
\draw [edge] (0,0) to (0,2);
\draw [edge] (1,0) to +(0,2);
\def\leftmost{-1.1}
\foreach \y/\l/\t in
{
1/{[a]}/{}}
{
\node [programlabel] (1) at (\leftmost, \y) {$\l$};
\draw [hor] (1.east -| 1.2,0) to (1.east);
\node [programlabel] at (\leftmost, \y-0.5) {\t};
}
\end{tz}
\end{calign}

\figurecaptionsuck
\caption{Notation for states of a groubit and a bit.}
\label{fig:states}
\figurecaptionpostsuck
\end{figure}
Just as classical bits are special cases of dits, so groubits are special cases of groudits.
\begin{definition}
\label{def:groubit}%
A \textit{groubit} $\B$ is the groudit with identity bijections, and with underlying groupoid $\cat{B}$ defined as follows, where $s,t$ are the source and target functions:
\begin{align}
\Ob(\cat{B}) &:= \Z_2
&
\Mor(\cat{B}) &:= \Z_2 \times \Z_2
&
s, t &:=  \Z_2 \times \Z_2 \stackrel {\pi_1} \to \Z_2
\end{align}
Composition is defined as follows: $(a,b) \circ (a,c) := (a,b \oplus c).$
\end{definition}

\noindent
So for $a,b \in \Z_2$, we write $(a,b)$ to denote a morphism of type $a \to a$. Using the terminology of Section~\ref{sec:overview}, we interpret $a$ as the logical bit, and $b$ as the internal bit. It follows from the composition law that the identity morphisms are of the form $(a,0)$. For the bijection data, we exploit the fact that the groupoid is in a natural way the disjoint union of two copies of $\Z_2$, and so the bijections have the type $\epsilon_i, \tau_i : G_i = \Z_2\to \Ob(B) = \Z_2$. We choose all 4 of these bijections to be the identity.

For every protocol we give in this paper, we describe an implementation for an arbitrary groudit, and prove correctness at this general level. However, for informal discussions of groudit phenomena, and for the explicit traces of each protocol that we give throughout Section~\ref{sec:protocols}, we talk in terms of groubits.

\paragraph{Dits.}
We can also describe classical dits using groupoids.
\begin{definition}
A \textit{dit} $\mathcal{D}$ is a discrete skeletal groupoid $\cat{D}$ with $d$ morphisms.
\end{definition}

\noindent
We recall that a groupoid is discrete when every morphism is an identity. For a dit $\mathcal{D}$, we write $[i]$ to denote a morphism $i \in \Mor(\cat{D})$. An ordinary classical bit is a  dit with $d=2$.

\paragraph{States.}
A \textit{state} of a groudit or dit is a morphism in the corresponding groupoid. Our dynamics are nondeterministic, so after a protocol, the final state of a system is in general a multiset drawn from the set of states. We indicate these multisets with a sum notation, with coefficients drawn from~$\N$. 

In our graphical calculus, a groudit is represented by a blue region, and  a classical dit by a yellow region. To indicate the state of the system at a given time, we draw a horizontal dashed line, and write the state to the left; see \autoref{fig:states}.

\paragraph{Operations.}
In our graphical calculus we define \textit{atomic operations}, and also \textit{derived operations} which are built from the atomic operations. We summarize these here, and show explicitly how they act on groubits and bits. This notation is all that is required to follow the protocol traces illustrated in Section~\ref{sec:protocols}. In all our diagrams, time flows from bottom to top. All operations listed here map every input state to a nonempty multiset, meaning that they will not fail. That makes them suitable building blocks for a groudit programming language.

\begin{figure}[b!]
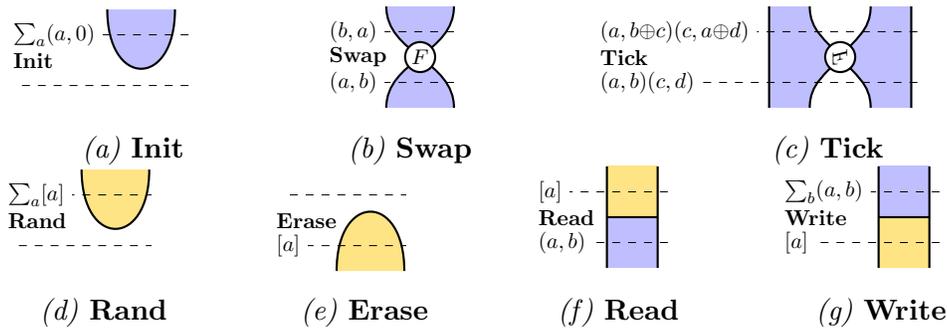

\figuretopsuck
\begin{calign}
\nonumber
\begin{tz}[scale=0.9]
\clip (-1.4,0) rectangle (1.22,-1.5);
\draw [surface, edge] (0,0) to [out=down, in=down, looseness=3] (1,0);
\def\leftmost{-1.5cm}
\foreach \y/\l/\t in
{-1.125/{}/{},
-0.375/{\sum_{a}(a,0)}/{\Init}}
{
\node [programlabel] (1) at (\leftmost, \y) {$\l$};
\draw [hor] (1.east -| 1.2,0) to (1.east);
\node [programlabel] at (\leftmost+0.0cm, \y-0.375) {\t};
}
\end{tz}
&\hspace{-2pt}
\begin{tz}[scale=0.9]
\draw [surface] (0,0) to [out=up, in=down] (1,1.5) to (0,1.5) to [out=down, in=up] (1,0);
\draw [edge] (0,0) to [out=up, in=down] (1,1.5);
\draw [edge] (1,0) to [out=up, in=down] (0,1.5);
\node [blob] at (0.5,0.75) {$F$};
\def\leftmost{-0.95cm}
\foreach \y/\l/\t in
{0.375/{(a,b)}/{},
1.125/{(b,a)}/{\Swap}}
{
\node [programlabel] (1) at (\leftmost, \y) {$\l$};
\draw [hor] (1.east -| 1.,0) to (1.east);
\node [programlabel] at (\leftmost+0.0cm, \y-0.375) {\t};
}
\end{tz}
&
\begin{tz}[scale=0.75,xscale=0.8,scale=0.9]
\path[surface,even odd rule] 
(0.25,0) to [out= up, in=-135]  (1,1) to [out= 45, in=down] (1.75,2) to (-0.75,2) to (-0.75,0) to (0.25,0)
(1.75,0) to [out= up, in=-45] (1,1) to [out= 135, in=down] (0.25,2) to (2.75,2) to (2.75,0) to (1.75,0);
\draw[edge] (0.25,0) to [out= up, in=-135]  (1,1) to[out=45, in=down] (1.75,2);
\draw[edge](1.75,0) to [out=up, in=-45] (1,1) to [out=135, in=down] (0.25,2);
\draw[edge] (-0.75,0) to +(0,2);
\draw[edge] (2.75,0) to +(0,2);
\node[blob,yscale=-1,rotate=90] at (1,1) {$F$};
\def\leftmost{-5.1cm}
\foreach \y/\l/\t in
{0.5/{(a,b) (c,d)}/{},
1.5/{(a,b\soplus c)(c,a\soplus d)}/{\Tick}}
{
\node [programlabel] (1) at (\leftmost, \y) {$\l$};
\draw [hor] (1.east -| 2.9,0) to (1.east);
\node [programlabel] at (\leftmost+0.0cm, \y-0.5) {\t};
}
\end{tz}
\\*\nonumber
\text{\hspace{0.85cm}\textit{(a)} \Init}
& \text{\hspace{0.5cm}\textit{(b)} \Swap}
& \text{\hspace{1.8cm} \textit{(c)} \Tick}
\end{calign}

\vspace{-20pt}
\begin{calign}
\nonumber
\begin{tz}[scale=0.9]
\clip (-1.2,0) rectangle (1.02,-1.5);
\draw [classical, edge] (0,0) to [out=down, in=down, looseness=3] (1,0);
\def\leftmost{-1.2cm}
\foreach \y/\l/\t in
{-1.125/{}/{},
-0.375/{\sum_{a}[a]}/{\Rand}}
{
\node [programlabel] (1) at (\leftmost, \y) {$\l$};
\draw [hor] (1.east -| 1.3,0) to (1.east);
\node [programlabel] at (\leftmost+0.0cm, \y-0.375) {\t};
}
\end{tz}
&\hspace{-10pt}
\begin{tz}[yscale=-1,scale=0.9]
\clip (-1.2,0) rectangle (1.02,-1.5);
\draw [classical, edge] (0,0) to [out=down, in=down, looseness=3] (1,0);
\def\leftmost{-1cm}
\foreach \y/\l/\t in
{-1.125/{}/{},
-0.375/{[a]}/{\Erase}}
{
\node [programlabel] (1) at (\leftmost, \y) {$\l$};
\draw [hor] (1.east -| 1.3,0) to (1.east);
\node [programlabel] at (\leftmost+0.0cm, \y-0.375) {\t};
}
\end{tz}
&
\begin{tz}[scale=0.75,yscale=1,scale=0.9]
\draw [surface] (0,0) rectangle +(1,1);
\draw [classical] (0,1) rectangle +(1,1);
\draw [edge] (0,0) to (0,2);
\draw [edge] (1,0) to +(0,2);
\draw [edge] (0,1) to +(1,0);
\def\leftmost{-1.5cm}
\foreach \y/\l/\t in
{0.5/{(a,b)}/{},
1.5/{[a]}/{\Read}}
{
\node [programlabel] (1) at (\leftmost, \y) {$\l$};
\draw [hor] (1.east -| 1.2,0) to (1.east);
\node [programlabel] at (\leftmost+0.0cm, \y-0.5) {\t};
}
\end{tz}
&
\begin{tz}[scale=0.75,scale=0.9]
\draw [classical] (0,0) rectangle +(1,1);
\draw [surface] (0,1) rectangle +(1,1);
\draw [edge] (0,0) to (0,2);
\draw [edge] (1,0) to +(0,2);
\draw [edge] (0,1) to +(1,0);
\def\leftmost{-2cm}
\foreach \y/\l/\t in
{0.5/{[a]}/{},
1.5/{\sum_b(a,b)}/{\Write}}
{
\node [programlabel] (1) at (\leftmost, \y) {$\l$};
\draw [hor] (1.east -| 1.2,0) to (1.east);
\node [programlabel] at (\leftmost+0.0cm, \y-0.5) {\t};
}
\end{tz}
\\*\nonumber
 \text{\hspace{0.75cm}\textit{(d)} \Rand}&
 \text{\hspace{0.55cm}\textit{(e)} \Erase}
 &
\text{\hspace{0.5cm}\textit{(f)} \Read}
& \text{\hspace{0.6cm}\textit{(g)} \Write}
\end{calign}

\figurecaptionsuck
\caption{Atomic groubit and bit operations.\label{fig:atomic}}
\end{figure}

\paragraph{Atomic operations.}
In \autoref{fig:atomic} we list the atomic operations involving a bit and a groubit. \autoref{fig:atomic}(a)--(c) shows the three groubit-only operations: \Swap and \Tick are deterministic, while \Init creates a groubit in a nondeterministic logical state. \autoref{fig:atomic}(c) uses a rotated and reflected letter to label the vertex, since it is represented algebraically by a rotated and reflected version of \autoref{fig:atomic}(b) under the dagger pivotal structure (see the discussion in Section~\ref{sec:groupoidsandactions} -- graphical calculus).

Note that the result of performing two successive \Tick operations between neighboring parties Alice and Bob, and Bob and Charlie, does not depend on the order of the operations; there is no race condition. Using the expression in \autoref{fig:atomic}(c) this becomes a simple isotopy, a crucial feature of our 2-dimensional graphical calculus.

\autoref{fig:atomic}(d)--(e) represents nondeterministic generation and erasure of a classical bit. \autoref{fig:atomic}(f)--(g) give the basic interactions between a groubit and a bit: \Read depicts the read-out of the logical state of a groubit, and \Write depicts the initialization of a groubit with given logical bit and random internal bit.

\begin{figure}[b!]
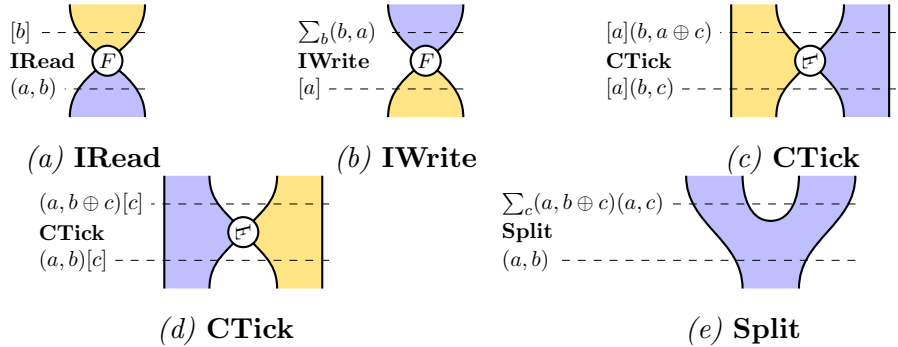

\begin{calign}
\nonumber
\begin{tz}
\begin{scope}
\clip (0,0) rectangle (1,0.75);
\draw [surface] (0,0) to [out=up, in=down] (1,1.5) to (0,1.5) to [out=down, in=up] (1,0);
\end{scope}
\begin{scope}
\clip (0,0.75) rectangle +(1,0.75);
\draw [classical] (0,0) to [out=up, in=down] (1,1.5) to (0,1.5) to [out=down, in=up] (1,0);
\end{scope}
\draw [edge] (0,0) to [out=up, in=down] (1,1.5);
\draw [edge] (1,0) to [out=up, in=down] (0,1.5);
\node [blob] at (0.5,0.75) {$F$};
\def\leftmost{-0.9cm}
\foreach \y/\l/\t in
{0.375/{(a,b)}/{},
1.125/{[b]}/{\IRead}}
{
\node [programlabel] (1) at (\leftmost, \y) {$\l$};
\draw [hor] (1.east -| 1.,0) to (1.east);
\node [programlabel] at (\leftmost+0.0cm, \y-0.375) {\t};
}
\end{tz}
&
\begin{tz}[yscale=1]
\begin{scope}
\clip (0,0) rectangle (1,0.75);
\draw [classical] (0,0) to [out=up, in=down] (1,1.5) to (0,1.5) to [out=down, in=up] (1,0);
\end{scope}
\begin{scope}
\clip (0,0.75) rectangle +(1,0.75);
\draw [surface] (0,0) to [out=up, in=down] (1,1.5) to (0,1.5) to [out=down, in=up] (1,0);
\end{scope}
\draw [edge] (0,0) to [out=up, in=down] (1,1.5);
\draw [edge] (1,0) to [out=up, in=down] (0,1.5);
\node [blob,yscale=1] at (0.5,0.75) {$F$};
\def\leftmost{-1.3cm}
\foreach \y/\l/\t in
{0.375/{[a]}/{},
1.125/{\sum_b (b,a)}/{\IWrite}}
{
\node [programlabel] (1) at (\leftmost, \y) {$\l$};
\draw [hor] (1.east -| 1.,0) to (1.east);
\node [programlabel] at (\leftmost+0.0cm, \y-0.375) {\t};
}
\end{tz}
&
\begin{tz}[scale=0.75,xscale=0.8]
\path[classical] 
(0.25,0) to [out= up, in=-135]  (1,1) to [out=135, in=down] (0.25,2) to (-0.75,2) to (-0.75,0);
\path[surface] (1.75,0) to [out=up, in=-45] (1,1) to [out= 45, in=down] (1.75,2) to (2.75,2) to (2.75,0);
\draw[edge] (0.25,0) to [out= up, in=-135]  (1,1) to[out=45, in=down] (1.75,2);
\draw[edge](1.75,0) to [out=up, in=-45] (1,1) to [out=135, in=down] (0.25,2);
\draw[edge] (-0.75,0) to +(0,2);
\draw[edge] (2.75,0) to +(0,2);
\node[blob,yscale=-1,rotate=90] at (1,1) {$F$};
\def\leftmost{-3.7cm}
\foreach \y/\l/\t in
{0.5/{[a] (b,c)}/{},
1.5/{[a](b,a\oplus c)}/{\CTick}}
{
\node [programlabel] (1) at (\leftmost, \y) {$\l$};
\draw [hor] (1.east -| 2.9,0) to (1.east);
\node [programlabel] at (\leftmost+0.0cm, \y-0.5) {\t};
}
\end{tz}
\\*\nonumber 
\text{\hspace{0.5cm}\textit{(a)} \IRead}
&\text{\hspace{0.75cm}\textit{(b)} \IWrite}
&\text{\hspace{1.2cm}\textit{(c)} \CTick}
\end{calign}

\vspace{-20pt}
\begin{calign}
\nonumber
\begin{tz}[scale=0.75,xscale=0.8]
\path[surface] 
(0.25,0) to [out= up, in=-135]  (1,1) to [out=135, in=down] (0.25,2) to (-0.75,2) to (-0.75,0);
\path[classical] (1.75,0) to [out=up, in=-45] (1,1) to [out= 45, in=down] (1.75,2) to (2.75,2) to (2.75,0);
\draw[edge] (0.25,0) to [out= up, in=-135]  (1,1) to[out=45, in=down] (1.75,2);
\draw[edge](1.75,0) to [out=up, in=-45] (1,1) to [out=135, in=down] (0.25,2);
\draw[edge] (-0.75,0) to +(0,2);
\draw[edge] (2.75,0) to +(0,2);
\node[blob,yscale=-1,rotate=90] at (1,1) {$F$};
\def\leftmost{-3.7cm}
\foreach \y/\l/\t in
{0.5/{(a,b)[c]}/{},
1.5/{(a,b\oplus c)[c]}/{\CTick}}
{
\node [programlabel] (1) at (\leftmost, \y) {$\l$};
\draw [hor] (1.east -| 2.9,0) to (1.east);
\node [programlabel] at (\leftmost+0.0cm, \y-0.5) {\t};
}
\end{tz}
&
\begin{tz}[scale=0.75]
\path[surface] 
(0,0) to [out=up, in=down] (-1,2) to (0,2) to [out=down,in=left] (0.5,1.2) to [out=right, in=down] (1,2) to (2,2) to [out=down, in=up] (1,0);
\draw[edge] (0,0) to [out=up, in=down] (-1,2);
\draw[edge] (0,2) to [out=down,in=left] (0.5,1.2) to [out=right, in=down] (1,2);
\draw[edge] (2,2) to [out=down, in=up] (1,0);
\def\leftmost{-4.4cm}
\foreach \y/\l/\t in
{0.5/{(a,b)}/{},
1.5/{\sum_c(a,b\oplus c) (a,c)}/{\Split}}
{
\node [programlabel] (1) at (\leftmost, \y) {$\l$};
\draw [hor] (1.east -| 2.,0) to (1.east);
\node [programlabel] at (\leftmost+0.0cm, \y-0.5) {\t};
}
\end{tz}
\\*\nonumber
\text{\hspace{1.2cm}\textit{(d)} \CTick}&
\text{\hspace{1.8cm}\textit{(e)} \Split}
\end{calign}

\figurecaptionsuck
\caption{Derived groubit and bit operations.\label{fig:derived}}
\end{figure}

\paragraph{Derived operations.}
In \autoref{fig:derived} we list the derived operations \IRead, \IWrite, \CTick and \Split. Note that \CTick comes in both left and right versions, distinguished by their images. We will see how they are defined in terms of atomic operations later in this section. 

\subsection{Graphical calculus}
\label{sec:groupoidsandactions}

\firstparagraph{Definition.}
Our graphical calculus represents groupoids, free actions and spans. We begin with an informal definition of the 2\-category formed by these structures. Throughout, we write `2-category' to refer to the weak structure, which is sometimes called `bicategory'. Similarly, we write `2-functor' to refer to a weak 2-functor, which is sometimes called a `pseudofunctor'.

\begin{definition}
\label{def:firstdefgpdact}
The 2\-category $\GAf$ is built from the following structures:
\begin{itemize}
\item \textit{objects} are finite skeletal groupoids $\cat G$, $\cat H$, ...;
\item a \textit{morphism} $S: \cat G \hto \cat H$ comprises, for any $a \in \Ob(\cat G)$ and $b \in \Ob(\cat H)$, a finite set $S_{a,b}$ equipped with commuting free\footnote{An action of a group $H$ on a set $X$ is \emph{free}, if for any $x\in X$, $h.x=x$ implies that $h=e_H$.} left- and right-actions of $\Aut_\cat G(a)$ and $\Aut_\cat H(b)$ respectively;
\item for morphisms $S,T: \cat G \hto \cat H$, a \textit{2-morphism} $\sigma: S \hTo T$ is an \textit{equivariant span}, comprising for all $a \in \Ob(\cat G)$ and $b \in  \Ob(\cat H)$ a function $\sigma_{a,b}: S_{a,b} \times T_{a,b} \to \N$, such that for all $g \in \cat \Aut_\cat G(a)$, $h \in \Aut_\cat H(b)$, $s \in S_{a,b}$ and $t \in T_{a,b}$ we have $\sigma_{a,b}(s,t) = \sigma_{a,b}(g.s.h, g.t.h).$
\end{itemize}
For the definition of composition, see Section~\ref{fulldefinition}.
\end{definition}

\noindent
Here $g.s.h \in S_{a,b}$ denotes the action on $s$ by $g$ on the left and $h$  on the right; since these actions commute, this is well-defined. Note the requirement that these left- and right-actions are free, which is important for guaranteeing that our constructions are well-defined. In the main part of this paper we will work with these structures informally; we give a formal 2\-categorical analysis in~\autoref{sec:bicategories}. 

\begin{definition}
\label{def:spandagger}
For an equivariant span $\sigma: S \hTo T$, we define its \textit{dagger} $\sigma ^\dag: T \hTo S$ as the converse: $\sigma^\dag_{a,b}(t,s) := \sigma_{a,b}(s,t)$.
\end{definition}

\begin{figure}[b]
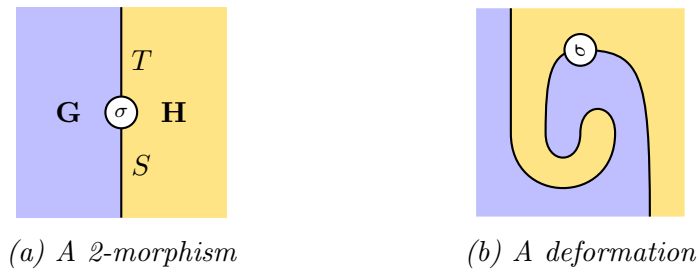

\figuretopsuck
\begin{calign}
\nonumber
\begin{tz}[scale=0.7]
\draw [surface] (0,0) rectangle (2,4);
\draw [classical] (2,0) rectangle +(2,4);
\node [blob, minimum width=12pt] at (2,2) { $\sigma$};
\draw [edge] (2,0) to (2,4);
\node at (1,2) {$\cat G$};
\node at (3,2) {$\cat H$};
\node [right] at (2,1) {$S$};
\node [right] at (2,3) {$T$};
\end{tz}
&
\begin{tz}[scale=0.66,scale=0.7, yscale=6/5]
\draw [surface] (1,5) to (1,2) to [out=down, in=down, looseness=1.5] (4,2) to [out=up, in=up, looseness=2] (3,2) to [out=down, in=down, looseness=2] (2,2) to [out=up, in=left, out looseness=1.5] (3,4) node [blob, rotate=90, minimum width=12pt] {$\sigma$} to [out=right, in=up, in looseness=2] (5,0) to (0,0) to (0,5);
\draw [classical] (1,5) to (1,2) to [out=down, in=down, looseness=1.5] (4,2) to [out=up, in=up, looseness=2] (3,2) to [out=down, in=down, looseness=2] (2,2) to [out=up, in=left, out looseness=1.5] (3,4) to [out=right, in=up, in looseness=2] (5,0) to (6,0) to (6,5);
\draw [edge] (1,5) to (1,2) to [out=down, in=down, looseness=1.5] (4,2) to [out=up, in=up, looseness=2] (3,2) to [out=down, in=down, looseness=2] (2,2) to [out=up, in=left, out looseness=1.5] (3,4) to [out=right, in=up, in looseness=2] (5,0);
\end{tz}
\\\nonumber
\textit{(a) A 2-morphism} & \textit{(b) A deformation}
\end{calign}

\figurecaptionsuck
\caption{Examples of the graphical calculus.}
\label{fig:graphicalcalculusexample}
\figurecaptionpostsuck
\end{figure}

\firstparagraph{Graphical calculus.}
We use a 2\-dimensional graphical calculus (see \autoref{fig:graphicalcalculusexample}(a)) to denote a 2\-morphism $\sigma$ in \GAf. This is the standard graphical calculus for 2\-categories~\cite{Selinger:2010}: objects \cat G, \cat H label the regions, morphisms $S,T: \cat G \hto \cat H$ label the wires, and 2\-morphisms $\sigma: S \hTo T$ label the vertices. We often drop the labels; also, white regions will always correspond to the trivial groupoid \cat 1 with one morphism. 
Stacking these pictures vertically performs composition of spans, stacking them horizontally performs bimodule composition, and reflecting them about a horizontal axis corresponds to the dagger operation of Definition~\ref{def:spandagger}. In fact, \GAf is a \textit{dagger pivotal 2\-category}~\cite[Section 2.1]{Carqueville:2016}, giving immense freedom in the graphical calculus: one may reflect, rotate and deform the pictures arbitrarily (holding the boundaries fixed), and these manipulations preserve equality of the diagram. For example, since the images of \autoref{fig:graphicalcalculusexample}(a) and (b) are deformations of each other with constant boundary, they represent equal 2\-morphisms.

\paragraph{Boundaries.} For every shaded region labelled by a skeletal groupoid \G, we have canonical boundaries drawn as follows:
\begin{calign}
\label{eq:boundaries}
\begin{tz}
\draw [surface] (0,0) rectangle (1,1.5);
\draw [edge] (0,0) to (0,1.5);
\end{tz}
&
\begin{tz}[xscale=-1]
\draw [surface] (0,0) rectangle (1,1.5);
\draw [edge] (0,0) to (0,1.5);
\end{tz}
\\*
\nonumber
L^\cat G: \cat 1 \hto \cat G & R^\cat G : \cat G \hto \cat 1 
\end{calign}
We define these as the following sets with free right (or left) $\Aut_\cat G(a)$-action, for all objects $a \in \cat G$:
\begin{align}
&&L^\cat G _{\bullet, a} &:= \Aut_\cat G(a) & (*, g, g') &\mapsto gg'
\\*
&&R^\cat G _{a, \bullet} &:= \Aut_\cat G (a) & (g', g, *) &\mapsto g'g
\end{align}
Here, $\bullet$ and $*$ denote the unique object and morphism of the terminal groupoid $\cat 1$, respectively.
That is, these boundaries are defined as the groupoid acting on itself, by left or right action. Using the pivotal structure, these boundaries give rise to the operations \Init, \Erase and \Split as presented in Section~\ref{sec:grouditsanddits}.

\subsection{Biunitaries}
\label{sec:biunitaries}

\noindent
Biunitaries are important structures from the theory of planar algebras (see Section~\ref{sec:relatedwork}) which play an essential role in our calculus.

\begin{figure*}[t]
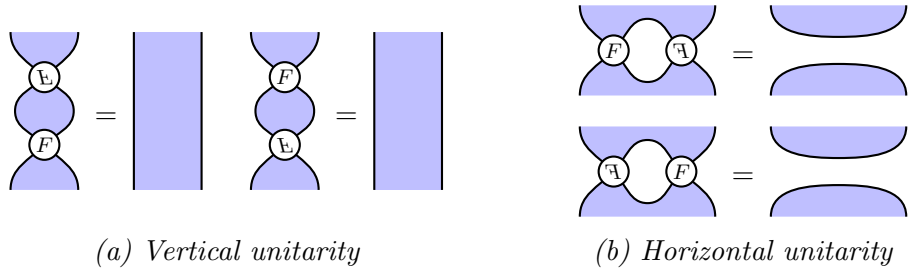

\figuretopsuck
\def\bigangle{150}
\def\smallangle{30}
\def \sidew {0.5}
\def \scl{0.6}
\begin{calign}
\nonumber
\begin{tz}[string,scale=\scl]
\path[surface,edge] (0.25,0) to [out=90, in=-135] (1,1) to [out=-45, in=90] (1.75,0);
\path[surface,edge] (0.25,3.5) to [out=-90, in=135] (1,2.5) to [out=45, in=-90] (1.75,3.5);
\path[surface,edge] (1,2.5) to [out=-135, in=90] (0.35, 1.75) to [out=-90, in=135] (1,1) to [out=45, in=-90] (1.65,1.75) to [out=90, in=-45] (1,2.5);
\node[blob] at (1,1) {$\F$};
\node[blob,yscale=-1] at (1,2.5) {$\F$};
\end{tz}
\eqgap=\eqgap
\begin{tz}[edge,scale=\scl]
\path[surface] (0.25,0) rectangle (1.75,3.5);
\draw (0.25,0) to (0.25,3.5);
\draw (1.75,0) to (1.75,3.5);
\end{tz}
\hspace{0.5cm}
\begin{tz}[string,scale=\scl]
\path[surface,edge] (0.25,0) to [out=90, in=-135] (1,1) to [out=-45, in=90] (1.75,0);
\path[surface,edge] (0.25,3.5) to [out=-90, in=135] (1,2.5) to [out=45, in=-90] (1.75,3.5);
\path[surface,edge] (1,2.5) to [out=-135, in=90] (0.35, 1.75) to [out=-90, in=135] (1,1) to [out=45, in=-90] (1.65,1.75) to [out=90, in=-45] (1,2.5);
\node[blob,yscale=-1] at (1,1) {\F};
\node[blob] at (1,2.5) {\F};
\end{tz}
\eqgap=\eqgap
\begin{tz}[edge,scale=\scl]
\path[surface] (0.25,0) rectangle (1.75,3.5);
\draw (0.25,0) to (0.25,3.5);
\draw (1.75,0) to (1.75,3.5);
\end{tz}
&
\vc{
$\begin{tz}[edge,scale=\scl]
\path[surface,draw] (3.25,0) to [out=90, in=-45] (2.5,1)to [out=-135, in=0] (1.75,0.3)  to [out=180, in=-45] (1,1) to [out=-135, in=90] (0.25,0);
\path[surface,draw] (0.25,2) to [out=-90, in=135] (1,1) to [out=45, in=180] (1.75,1.7) to [out=0, in=135] (2.5,1) to [out=45, in=-90] (3.25,2);
\node[blob] at (1,1) {\F};
\node[blob,yscale=-1,rotate=180]at (2.5,1) {\F};
\end{tz}
\eqgap =\eqgap
\begin{tz}[edge,scale=\scl] 
\path[surface,edge] (3.25,2) to [out=-90, in=0] (1.75,1.3)  to [out=180, in=-90] (0.25,2);
\path[surface,edge] (3.25,0) to [out=90, in=0] (1.75,0.7)  to [out=180, in=90] (0.25,0);
\end{tz}$
\\[20pt]
$\begin{tz}[edge,scale=\scl]
\path[surface,edge] (3.25,0) to [out=90, in=-45] (2.5,1)to [out=-135, in=0] (1.75,0.3)  to [out=180, in=-45] (1,1) to [out=-135, in=90] (0.25,0);
\path[surface,edge] (0.25,2) to [out=-90, in=135] (1,1) to [out=45, in=180] (1.75,1.7) to [out=0, in=135] (2.5,1) to [out=45, in=-90] (3.25,2);
\node[blob,yscale=-1,rotate=180] at (1,1) {\F};
\node[blob] at (2.5,1) {\F};
\end{tz}
\eqgap =\eqgap
\begin{tz} [edge,scale=\scl]
\path[surface,edge] (3.25,2) to [out=-90, in=0] (1.75,1.3)  to [out=180, in=-90] (0.25,2);
\path[surface,edge] (3.25,0) to [out=90, in=0] (1.75,0.7)  to [out=180, in=90] (0.25,0);
\end{tz}$
}
\\\nonumber
\text{\em(a) Vertical unitarity}
&
\text{\em(b) Horizontal unitarity}
\end{calign}

\figurecaptionsuck
\caption{\label{fig:biunitarity}The biunitarity equations.}
\figurecaptionpostsuck
\end{figure*}

\begin{definition}
In \GAf, a biunitary on a skeletal groupoid \cat G is a unitary 2-morphism
\begin{equation}\label{eq:biunitary}
\begin{tz}[scale=0.9]
\draw [surface] (0,0) to [out=up, in=down] (1,1.5) to (0,1.5) to [out=down, in=up] (1,0);
\draw [edge] (0,0) to [out=up, in=down] (1,1.5);
\draw [edge] (1,0) to [out=up, in=down] (0,1.5);
\node [blob] at (0.5,0.75) {$F$};
\end{tz}\end{equation} fulfilling the equations depicted in \autoref{fig:biunitarity}.
\end{definition}

\noindent
The source and target of the biunitary is the composite 1-morphism $R^{\cat{G}}\circ L^{\cat{G}}: \cat{1}\hto \cat{1}$ which evaluates to the set $\Mor(\cat G)$ of morphisms of the skeletal groupoid. Concretely, therefore, a biunitary is an automorphism of $\Mor(\cat G)$ satisfying an algebraic condition; it plays the role of a generalized \Swap map in our groudit programming language. The following theorem determines this condition precisely.

\begin{theorem} A biunitary on a skeletal groupoid $\G$ is a bijection $F: \morG \to \morG$ such that for all \mbox{$a,b \in \obG$}, we have:
\begin{equation}
\label{eq:biunitarityalgebraically}
\left| F(\Aut_\cat G(a)) \cap \Aut _\cat G(b)\right| = 1
\end{equation}
\end{theorem}
\begin{proof} The equations of \autoref{fig:biunitarity}(a) say that $F$ is unitary, which means precisely that it acts as a permutation on $\morG$. The equations of  \autoref{fig:biunitarity}(b) are equivalent to the composite of \autoref{fig:biunitarycalculation} being the identity.
\begin{figure}[b]
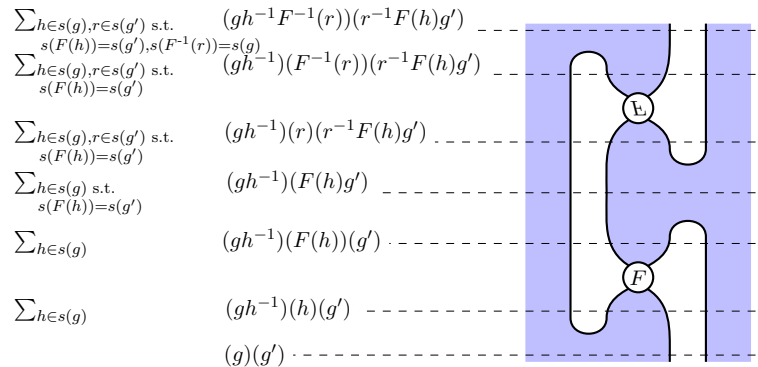

\figuretopsuck
\def\sideangle{30}
\def\nwangle{180-\sideangle}
\def\neangle{\sideangle}
\def\swangle{180+\sideangle}
\def\seangle{-\sideangle}
\[
\hspace{2cm}
\begin{tz}[scale=0.6, yscale=0.75]
\path[surface] (2.5, 0)
to (2.5, 5)
to (1.5,5)
to (1.5, 1.3)
to [out=down, in=down, looseness=2](0.7, 1.3)
to [out=up, in=\seangle](0, 2.3)
to [out=\swangle, in=up](-0.7,0)
to [out=down, in=\nwangle](-0,-2.3)
to [out=\neangle, in=down](0.7, -1.3)
to [out=up, in=up, looseness=2](1.5,-1.3)
to (1.5, -5)
to (2.5, -5);
\draw [edge](1.5,5)
to (1.5, 1.3)
to [out=down, in=down, looseness=2](0.7, 1.3)
to [out=up, in=\seangle](0, 2.3)
to  [out=\swangle, in=up](-0.7, 0)
to [out=down, in=\nwangle](-0,-2.3)
to [out=\neangle, in=down](0.7, -1.3)
to [out=up, in=up, looseness=2](1.5,-1.3)
to (1.5, -5);
\draw [surface] (-2.5,0)
to (-2.5, 5)
to (0.7, 5)
to (0.7, 4.5)
to [out=down, in=\neangle](0, 2.7)
to [out=\nwangle, in=down](-0.7, 3.7)
to [out=up, in=up, looseness=2](-1.5, 3.7)
to (-1.5, 0)
to (-1.5, -3.7)
to [out=down, in=down, looseness=2](-0.7, -3.7)
to [out=up, in=\swangle](0,-2.7)
to [out=\seangle, in=up](0.7, -4.5)
to (0.7, -5)
to (-2.5,-5);
\draw [edge](0.7, 5)
to (0.7, 4.5)
to [out=down, in=\neangle](0, 2.7)
to [out=\nwangle, in=down](-0.7, 3.7)
to [out=up, in=up, looseness=2](-1.5, 3.7)
to(-1.5, 0)
to (-1.5, -3.7)
to [out=down, in=down, looseness=2](-0.7, -3.7)
to [out=up, in=\swangle](0,-2.7)
to [out=\seangle, in=up](0.7, -4.5)
to (0.7, -5); 
\node[blob] at (0,-2.5) {\F};
\node[blob,yscale=-1] at (0,2.5) {\F};
\def\leftmost{-14cm}
\def\sp{\phantom{\hspace{0.8cm}}}
\foreach \y/\l/\t in
{-4.8/{\phantom{\hspace{2.6cm}}\,\,\,\,\,\sp (g)(g')}/{},
-3.5/{\sum_{h\in s(g)}\phantom{\hspace{1.5cm}}\,\sp (gh^{-1})(h)(g')}/{},
-1.5/{\sum_{h\in s(g)}\phantom{\hspace{1.5cm}}\sp (gh^{-1})(F(h))(g')}/{},
0/{\sum_{\substack{h\in s(g)\text{ s.t.}\phantom{\hspace{0.5cm}}\\ s(F(h))=s(g')}}\,\,\,\,\,\,\,\,\,\sp (gh^{-1})(F(h)g')}/{},
1.5/{\sum_{\substack{h\in s(g), r\in s(g')\text{ s.t.}\\s(F(h))=s(g')\phantom{\hspace{0.5cm}}}}\sp (gh^{-1})(r)(r^{-1}F(h)g')}/{},
3.5/{\sum_{\substack{h\in s(g), r\in s(g')\text{ s.t.}\\s(F(h))=s(g')\phantom{\hspace{0.5cm}}}}\hspace{-0.1cm}\,\sp (gh^{-1})(F^{-1}(r))(r^{-1}F(h)g')}/{},
4.8/{\sum_{\substack{h\in s(g), r\in s(g')\text{ s.t.}\phantom{\hspace{1.6cm}}\\s(F(h))=s(g'), s(F^{\text-1}(r))=s(g)}}\hspace{-1.7cm}\,\sp (gh^{-1}F^{-1}(r))(r^{-1}F(h)g')}/{}
}
{
\node [programlabel] (1) at (\leftmost, \y) {$\l$};
\draw [hor] (1.east -| 2.6,0) to (1.east);
\node [programlabel] at (\leftmost+0.0cm, \y-0.5) {\t};
}
\end{tz}
\]

\figurecaptionsuck
\caption{Verifying the action of a biunitary.}
\label{fig:biunitarycalculation}
\end{figure}
This holds just when, for all $a,b \in \Ob(\cat G)$ and for all $g\in \Aut_\cat G(a)$ and $g'\in \Aut_\G(b)$, there are unique $h\in \Aut_\G(a)$, $r\in \Aut_\G(b)$ with $s(F(h)) =b$ and $s(F^{-1}(r))=a$ satisfying the following conditions:
\begin{calign}
\nonumber
gh^{-1}F^{-1}(r) = g
&
r^{-1} F(h) g'=g'
\end{calign}
In other words for any two objects $a,b$ in the groupoid there is a unique pair $(h,r)\in \Aut_\G(a)\times \Aut_\G(b)$ such that $F(h)=r$. More concisely, $\left|F(\Aut_\G(a))\cap \Aut_\G(b)\right|=~\!1$.
\end{proof} 

\paragraph{Classification.} We now classify biunitaries in terms of groudits. This shows that biunitaries are tractable algebraic objects.

\begin{theorem}
\label{thm:biunitaryclassification}
For a skeletal groupoid \cat G, groudits on \cat G are in bijective correspondence with biunitaries on \cat G.
\end{theorem}
\begin{proof}
Define a  \emph{balancer} $\epsilon$ for \cat G to be a choice for all objects $a \in \Ob(\cat G)$ of a bijection \mbox{$\epsilon_a: \Aut_\cat G (a) \to \Ob(\cat G)$}. Clearly for any $b \in \Ob(\cat G)$ we have
\begin{equation}
\label{eq:balancersourcecondition}
s(\epsilon_a^{-1}(b)) = a.
\end{equation}
It is easy to see from the definition that a groudit is precisely a skeletal groupoid equipped with a pair of balancers. Given a balancer $\epsilon$, we define functions $\epsilon_1, \epsilon_2$ as follows:
\def\hsuck{\hspace{-6pt}}
\begin{align}
\hsuck\epsilon_1 &: \morG \to \obG {\times} \obG
&
&\epsilon_1(g) := (sg, \epsilon_{sg}(g))
\\
\hsuck\epsilon_2 &: \obG {\times} \obG \to \morG
&
&\epsilon_2(a,b) := \epsilon_a^{-1}(b)
\end{align}
We can show that $\epsilon_1$ and $\epsilon_2$ are inverse:
\begin{align*}
&\epsilon_1(\epsilon_2(a,b)) = (s(\epsilon_a^{-1}(b)), \epsilon_{s(\epsilon_a^{-1}(b))}(\epsilon_a^{-1}(b))
\superequals{eq:balancersourcecondition} (a, \epsilon_a(\epsilon_a ^{-1}(b))) = (a,b)
\\[5pt]
&\epsilon_2(\epsilon_1(g)) = \epsilon_{s(g)}^{-1}(\epsilon_{s(g)} ^{}(g))= g
\end{align*}
We now give the first direction of the main bijective correspondence. Suppose  $\epsilon, \tau$ are balancers for \cat G. Then we define a biunitary \mbox{$F_{\epsilon, \tau} : \morG \to \morG$} as the following composite, where $\gamma$ is the swap map for the cartesian product:
\begin{align*}
\label{eq:associatedbiunitary}
&\morG
\stackrel {\epsilon_1} \to
\obG {\times} \obG \stackrel \gamma \to \obG {\times} \obG
\stackrel {\tau_2} \to \morG
\end{align*}
Then a simple calculation shows the following:
\begin{align}
F_{\epsilon, \tau} &(g) = \tau ^{-1} _{\epsilon _{s(g)} (g)}(s(g)) \in \Aut_\cat G(\epsilon_{s(g)}(g))
\end{align}
By construction, $F_{\epsilon,\tau}$ is unitary, since it is a composite of bijections. To show it is biunitary, suppose now that $g,g'\in \Aut_\G(a)$ such that $s(F_{\epsilon,\tau}(g))=s(F_{\epsilon,\tau}(g'))$. Then by equation~\eqref{eq:balancersourcecondition}, we have $\epsilon_{s(g)}(g) = \epsilon _{s(g')}(g')$, and since $s(g) = s(g') =a$ we therefore have $\epsilon_a(g) = \epsilon_a(g')$, and since $\epsilon_a$ is a bijection we have $g=g'$.
\def\scl{0.8}
\begin{figure}[b!]
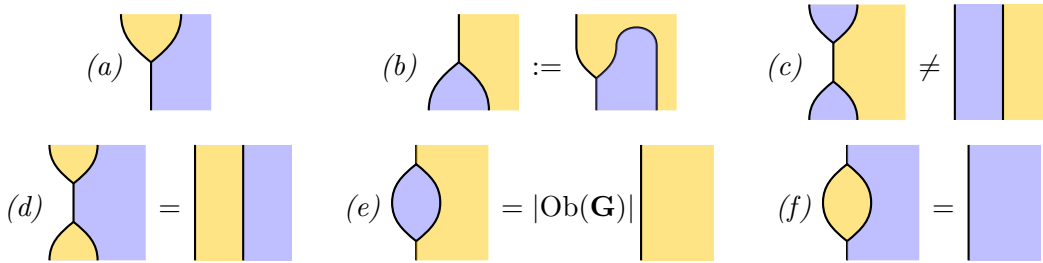

\tikzset{every picture/.style={scale=0.8, yscale=0.8}}
\figuretopsuck
\begin{calign}
\nonumber
\text{\textit{(a)}}
\hspace{-0.5cm}
\begin{tz}
\path [use as bounding box] (-1,0) rectangle +(2,2);
\draw [surface] (0,0) to (0,1) to [out=45, in=down] (0.5,2) to (1,2) to (1,0);
\draw [classical] (0,1) to [out=135, in=down] (-0.5,2) to (0.5,2) to [out=down, in=45] (0,1);
\draw [edge] (0,0) to (0,1) to [out=135, in=down] (-0.5,2);
\draw [edge] (0,1) to [out=45, in=down] (0.5,2);
\end{tz}
&
\text{\textit{(b)}}\,
\begin{tz}[xscale=1, yscale=-1]
\draw [classical] (0,0) to (0,1) to [out=45, in=down] (0.5,2) to (1,2) to (1,0);
\draw [surface] (0,1) to [out=135, in=down] (-0.5,2) to (0.5,2) to [out=down, in=45] (0,1);
\draw [edge] (0,0) to (0,1) to [out=135, in=down] (-0.5,2);
\draw [edge] (0,1) to [out=45, in=down] (0.5,2);
\end{tz}
:=
\begin{tz}[scale=0.6666]
\draw [classical] (-0.5,3) to (-0.5,2) to [out=down, in=135] (0,1) to (0,0) to (2,0) to (2,3);
\draw [edge, , fill=white, postaction={surface}] (0,0) to (0,1) to [out=45, in=down] (0.5,2) to [out=up, in=up, looseness=2] (1.5,2) to (1.5,0);
\draw [edge] (-0.5,3) to (-0.5,2) to [out=down, in=135] (0,1);
\end{tz}
&
\text{\textit{(c)}}
\begin{tz}[scale=\scl]
\draw [classical] (0,0) to [out=up, in=-135] (0.5,1) to (0.5,2) to [out=135, in=down] (0,3) to (2,3) to (2,0);
\path [fill=white, postaction=surface] (1,0) to [out=up, in=-45] (0.5,1) to (0.5,2) to [out=45, in=down] (1,3) to (0,3) to [out=down, in=135] (0.5,2) to (0.5,1) to [out=-135, in=up] (0,0);
\draw [edge] (1,0) to [out=up, in=-45] (0.5,1) to (0.5,2) to [out=45, in=down] (1,3);
\draw [edge] (0,3) to [out=down, in=135] (0.5,2);
\draw [edge] (0,0) to [out=up, in=-135] (0.5,1);
\end{tz}
\neq
\begin{tz}[scale=\scl]
\draw [surface] (0,0) rectangle +(1,3);
\draw [classical] (1,0) rectangle +(1,3);
\draw [edge] (0,0) to +(0,3);
\draw [edge] (1,0) to +(0,3);
\end{tz}
\\
\nonumber
\text{\textit{(d)}}
\begin{tz}[scale=\scl]
\draw [surface] (0,0) to [out=up, in=-135] (0.5,1) to (0.5,2) to [out=135, in=down] (0,3) to (2,3) to (2,0);
\path [fill=white, postaction=classical] (1,0) to [out=up, in=-45] (0.5,1) to (0.5,2) to [out=45, in=down] (1,3) to (0,3) to [out=down, in=135] (0.5,2) to (0.5,1) to [out=-135, in=up] (0,0);
\draw [edge] (1,0) to [out=up, in=-45] (0.5,1) to (0.5,2) to [out=45, in=down] (1,3);
\draw [edge] (0,3) to [out=down, in=135] (0.5,2);
\draw [edge] (0,0) to [out=up, in=-135] (0.5,1);
\end{tz}
=
\begin{tz}[scale=\scl]
\draw [classical] (0,0) rectangle +(1,3);
\draw [surface] (1,0) rectangle +(1,3);
\draw [edge] (0,0) to +(0,3);
\draw [edge] (1,0) to +(0,3);
\end{tz}
&
\text{\textit{(e)}}\,
\begin{tz}[scale=\scl]
\draw [edge] (0,0) to (0,3);
\draw [classical] (0,0) to (0,0.5) to [out=135, in=down] (-0.5,1.5) to [out=up, in=-135] (0,2.5) to (0,3) to (1.5,3) to (1.5,0);
\draw [fill=white, postaction={surface,edge}] (0,0.5) to [out=45, in=down] (0.5,1.5) to [out=up, in=-45] (0,2.5) to [out=-135, in=up] (-0.5,1.5) to [out=down, in=135] (0,0.5);
\end{tz}
=
|\Ob(\G)|
\begin{tz}[scale=\scl]
\draw [classical] (0,0) rectangle (1.5,3);
\draw [edge] (0,0) to (0,3);
\end{tz}
&\text{\textit{(f)}}\,
\begin{tz}[scale=\scl]
\draw [edge] (0,0) to (0,3);
\draw [surface] (0,0) to (0,0.5) to [out=135, in=down] (-0.5,1.5) to [out=up, in=-135] (0,2.5) to (0,3) to (1.5,3) to (1.5,0);
\draw [fill=white, postaction={classical,edge}] (0,0.5) to [out=45, in=down] (0.5,1.5) to [out=up, in=-45] (0,2.5) to [out=-135, in=up] (-0.5,1.5) to [out=down, in=135] (0,0.5);
\end{tz}
=
\begin{tz}[scale=\scl]
\draw [surface] (0,0) rectangle (1.5,3);
\draw [edge] (0,0) to (0,3);
\end{tz}
\end{calign}

\figurecaptionsuck
\caption{Building blocks for the measurement calculus.}
\figurecaptionpostsuck
\label{fig:measurement}
\end{figure}

We now give the reverse direction of the main bijective correspondence. Given a biunitary $F: \morG \to \morG$, we define balancers $\epsilon^F, \tau^F$ for all $a \in \Ob(\G)$ \mbox{and $g \in \Mor(\cat G)$ as}
\begin{align}
\epsilon^F_a(g) &:= s(F(g))
&
\tau^F_a(g) := s(F^{-1}(g))
\end{align}
We must show that for all $a \in \Ob(\G)$,  $\epsilon_a^F,\tau_a^F:\Aut _\G(a) \to \obG$ are bijections. First, surjectivity. For any $b\in \obG$, using the biunitarity property \eqref{eq:biunitarityalgebraically}, pick the unique morphism $g\in F(\Aut_\G(a))\cap \Aut_\G(b)$. Then $F^{-1}(g) \in \Aut_\G(a)$ and $b=s(g)=s(F(F^{-1}(g)))= \epsilon_a^F(F^{-1}g)$. A similar proof shows surjectivity of $\tau^F_G$. Next, injectivity. Suppose that $g,h\in \Aut_\G(a)$ with $\epsilon_a^F(g) = \epsilon_a^F(h)$; then $s(F(g))=s(F(h))$. Then $F(g),F(h)\in F(\Aut_\G(a)) \cap \Aut_\G(s(F(g)))$. Then by the biunitarity property \eqref{eq:biunitarityalgebraically}, we conclude that $F(g)=F(h)$ and therefore that $g=h$.

Finally, we show that the main correspondence is indeed bijective.
In one direction, for a pair of balancers $(\epsilon,\tau)$ with associated biunitary $F_{\epsilon,\tau}$ and $g\in \Aut_\cat G(a)$, then by \eqref{eq:balancersourcecondition} we have $\epsilon _a^{F_{\epsilon,\tau}}(g)=s(F_{\epsilon,\tau}(g)) = \epsilon_{a}(g)$ and similarly $\tau_a^{F_{\epsilon,\tau}}(g) =s(F_{\epsilon,\tau}^{-1}(g)) = \tau^F_a(g)$. In the other direction, given a biunitary $F$ and $g \in \Aut_\G(a)$, we observe that $F_{\epsilon^F,\tau^F}(g) = (\tau^F_{\epsilon^F_{a}(g)})^{-1}(a)$. To show that this equals $F(g)$, we have to show that $\tau^F_{\epsilon^F_a(g)}(F(g)) = a$. And indeed, we have $\tau^F_{\epsilon^F_G(g)}(F(g)) = s(F^{-1}(F(g)))=s(g)=a$.
\end{proof}

\subsection{Measurements}

\ignore{\firstparagraph{Syntax.}}
In Section~\ref{sec:grouditsanddits} we described classical dits using discrete groupoids. In the graphical calculus we draw them as yellow regions, to distinguish them from groudits which we draw in blue. There is an important difference: while blue regions are equipped with a biunitary of the form \eqref{eq:biunitary}, yellow regions are not equipped with any such structure.
\begin{figure}[b!]
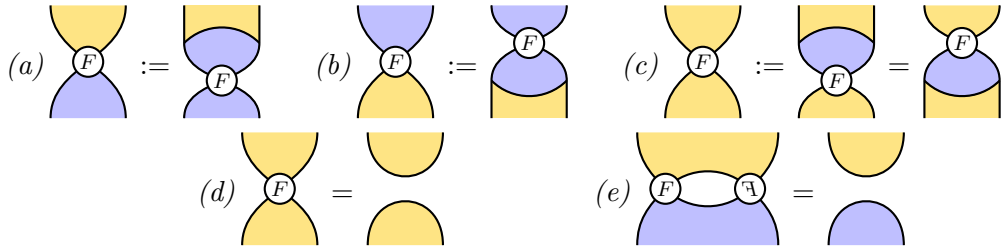

\figuretopsuck
\begin{calign}
&
\nonumber
\text{\textit{(a)}}
\begin{tz}
\begin{scope}
\clip (0,0) rectangle (1,0.75);
\draw [surface] (0,0) to [out=up, in=down] (1,1.5) to (0,1.5) to [out=down, in=up] (1,0);
\end{scope}
\begin{scope}
\clip (0,0.75) rectangle +(1,0.75);
\draw [classical] (0,0) to [out=up, in=down] (1,1.5) to (0,1.5) to [out=down, in=up] (1,0);
\end{scope}
\draw [edge] (0,0) to [out=up, in=down] (1,1.5);
\draw [edge] (1,0) to [out=up, in=down] (0,1.5);
\node [blob] at (0.5,0.75) {$F$};
\end{tz}
:=
\begin{tz}
\draw [classical] (0,1.5) rectangle (1,1);
\draw [fill=white, postaction={surface}] (0,0) to [out=up, in=down] (1,1) to [out=135, in=45] (0,1) to [out=down, in=up] (1,0);
\draw [edge] (1,0) to [out=up, in=down] (0,1) to +(0,0.5);
\draw [edge] (0,0) to [out=up, in=down] (1,1) to +(0,0.5);
\draw [edge] (0,1) to [out=45, in=135] (1,1);
\node [blob] at (0.5,0.5) {$F$};
\end{tz}
&
\text{\textit{(b)}}
\begin{tz}[yscale=-1]
\begin{scope}
\clip (0,0) rectangle (1,0.75);
\draw [surface] (0,0) to [out=up, in=down] (1,1.5) to (0,1.5) to [out=down, in=up] (1,0);
\end{scope}
\begin{scope}
\clip (0,0.75) rectangle +(1,0.75);
\draw [classical] (0,0) to [out=up, in=down] (1,1.5) to (0,1.5) to [out=down, in=up] (1,0);
\end{scope}
\draw [edge] (0,0) to [out=up, in=down] (1,1.5);
\draw [edge] (1,0) to [out=up, in=down] (0,1.5);
\node [blob] at (0.5,0.75) {$F$};
\end{tz}
:=
\begin{tz}[yscale=-1]
\draw [classical] (0,1.5) rectangle (1,1);
\draw [fill=white, postaction={surface}] (0,0) to [out=up, in=down] (1,1) to [out=135, in=45] (0,1) to [out=down, in=up] (1,0);
\draw [edge] (1,0) to [out=up, in=down] (0,1) to +(0,0.5);
\draw [edge] (0,0) to [out=up, in=down] (1,1) to +(0,0.5);
\draw [edge] (0,1) to [out=45, in=135] (1,1);
\node [blob] at (0.5,0.5) {$F$};
\end{tz}
&
\text{\textit{(c)}}
\begin{tz}
\draw [classical] (0,0) to [out=up, in=down] (1,1.5) to (0,1.5) to [out=down, in=up] (1,0);
\draw [edge] (0,0) to [out=up, in=down] (1,1.5);
\draw [edge] (1,0) to [out=up, in=down] (0,1.5);
\node [blob] at (0.5,0.75) {$F$};
\end{tz}
:=
\begin{tz}
\draw [classical] (0,1.5) rectangle (1,1);
\begin{scope}
\clip (0,0.5) rectangle (1,1.5);
\draw [fill=white, postaction={surface}] (0,0) to [out=up, in=down] (1,1) to [out=135, in=45] (0,1) to [out=down, in=up] (1,0);
\end{scope}
\begin{scope}
\clip (0,0) rectangle (1,0.5);
\draw [fill=white, postaction={classical}] (0,0) to [out=up, in=down] (1,1) to [out=135, in=45] (0,1) to [out=down, in=up] (1,0);
\end{scope}
\draw [edge] (1,0) to [out=up, in=down] (0,1) to +(0,0.5);
\draw [edge] (0,0) to [out=up, in=down] (1,1) to +(0,0.5);
\draw [edge] (0,1) to [out=45, in=135] (1,1);
\node [blob] at (0.5,0.5) {$F$};
\end{tz}
=
\begin{tz}[yscale=-1]
\draw [classical] (0,1.5) rectangle (1,1);
\begin{scope}
\clip (0,0.5) rectangle (1,1.5);
\draw [fill=white, postaction={surface}] (0,0) to [out=up, in=down] (1,1) to [out=135, in=45] (0,1) to [out=down, in=up] (1,0);
\end{scope}
\begin{scope}
\clip (0,0) rectangle (1,0.5);
\draw [fill=white, postaction={classical}] (0,0) to [out=up, in=down] (1,1) to [out=135, in=45] (0,1) to [out=down, in=up] (1,0);
\end{scope}
\draw [edge] (1,0) to [out=up, in=down] (0,1) to +(0,0.5);
\draw [edge] (0,0) to [out=up, in=down] (1,1) to +(0,0.5);
\draw [edge] (0,1) to [out=45, in=135] (1,1);
\node [blob] at (0.5,0.5) {$F$};
\end{tz}
\end{calign}

\vspace{-10pt}
\begin{calign}
\nonumber
&\text{\textit{(d)}}
\begin{tz}
\draw [classical] (0,0) to [out=up, in=down] (1,1.5) to (0,1.5) to [out=down, in=up] (1,0);
\draw [edge] (0,0) to [out=up, in=down] (1,1.5);
\draw [edge] (1,0) to [out=up, in=down] (0,1.5);
\node [blob] at (0.5,0.75) {$F$};
\end{tz}
=
\begin{tz}
\draw [classical, edge] (0,0) to [out=up, in=up, looseness=2] (1,0);
\draw [classical, edge] (0,1.5) to [out=down, in=down, looseness=2] +(1,0);
\end{tz}
&
\text{\textit{(e)}}
\begin{tz}[scale=0.75]
\draw [surface, edge] (0,0) to [out=up, in=-135] (0.5,1) to [out=-45, in=-135] (2,1) to [out=-45, in=up] (2.5,0);
\draw [classical, edge] (0,2) to [out=down, in=135] (0.5,1) to [out=45, in=135] (2,1) to [out=45, in=down] (2.5,2);
\node [blob] at (0.5,1) {$F$};
\node [blob, xscale=-1] at (2,1) {$F$};
\end{tz}
=
\begin{tz}
\draw [surface, edge] (0,0) to [out=up, in=up, looseness=2] (1,0);
\draw [classical, edge] (0,1.5) to [out=down, in=down, looseness=2] +(1,0);
\end{tz}
\end{calign}

\figurecaptionsuck
\caption{Yellow-blue and yellow-yellow versions of the biunitary.}
\figurecaptionpostsuck
\label{fig:redredbiunitary}
\end{figure}

\begin{figure*}
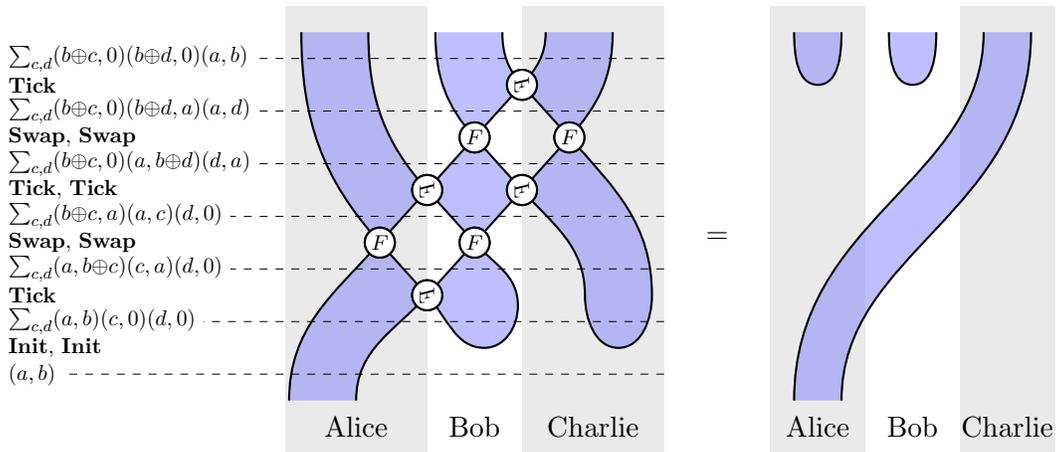

\[
\begin{tz}[protocoldiagram,xscale=0.9]
\path [background] (-1,-1) rectangle +(3,8.5);
\path [background] (4,-1) rectangle +(3,8.5);
\begin{scope}[scale=2]
\path[surface,even odd rule]
    (0.25,0) to [out=up, in=-135] (1,1) node[blob,rotate=-90,yscale=-1] {$F$} to [out=45, in=-135] (2.5,2.5) to [out=45, in=down] (2.25+\hrt,3.5) to (2.25,3.5) to [out=down, in=45] (2,3) to (0.5,1.5) node[blob] {$F$} to [out=-135, in=up] (0.25-\hrt,0)
    (2.5-4*\hrt,3.5) to [out=down, in=135]  (0.5,1.5) to (1,1) to [out=-45, in=left] (1.25+0.5*\hrt,0.5) to [out=right, in=down] (1.25+\hrt,0.9) to [out=up, in=-45] (1.5,1.5) node [blob] {$F$} to (1,2) node [blob,rotate=-90,yscale=-1] {$F$} to [out=135, in=down] (2.5-3*\hrt,3.5)
    (2.5-2*\hrt,3.5) to [out=down, in=135] (1.5,2.5) node[blob] {$F$} to (2,2) node [blob, rotate=-90, yscale=-1] {$F$} to [out=-45, in=up] (1.25+2*\hrt,1) to  [out=down, in=left] (1.25+2.5*\hrt,0.5) to [out=right, in=down] (1.25+3*\hrt,1) to [out=up, in=-45] (2.5,2.5) node [blob] {$F$} to (2,3) node[blob,rotate=-90,yscale=-1] {$F$} to [out=135, in=down] (2.5-\hrt,3.5)
    ;
\path[edge]
    (0.25,0) to [out=up, in=-135] (1,1) node[blob,rotate=-90,yscale=-1] {$F$} to [out=45, in=-135] (2.5,2.5) to [out=45, in=down] (2.25+\hrt,3.5) (2.25,3.5) to [out=down, in=45] (2,3) to (0.5,1.5) node[blob] {$F$} to [out=-135, in=up] (0.25-\hrt,0)
    (2.5-4*\hrt,3.5) to [out=down, in=135]  (0.5,1.5) to (1,1) to [out=-45, in=left] (1.25+0.5*\hrt,0.5) to [out=right, in=down] (1.25+\hrt,0.9) to [out=up, in=-45] (1.5,1.5) node [blob] {$F$} to (1,2) node [blob,rotate=-90,yscale=-1] {$F$} to [out=135, in=down] (2.5-3*\hrt,3.5)
    (2.5-2*\hrt,3.5) to [out=down, in=135] (1.5,2.5) node[blob] {$F$} to (2,2) node [blob, rotate=-90, yscale=-1] {$F$} to [out=-45, in=up] (1.25+2*\hrt,1) to  [out=down, in=left] (1.25+2.5*\hrt,0.5) to [out=right, in=down] (1.25+3*\hrt,1) to [out=up, in=-45] (2.5,2.5) node [blob] {$F$} to (2,3) node[blob,rotate=-90,yscale=-1] {$F$} to [out=135, in=down] (2.5-\hrt,3.5)
    ;
\end{scope}
\begin{scope}[xshift=-7cm, yshift=0.5cm]
\foreach \y/\l/\t in
{0/{(a,b)}/{},
1/{\sum_{c,d}(a,b) (c,0) (d,0)}/{\Init, \Init},
2/{\sum_{c,d}(a,b\soplus c)(c,a)(d,0)}/{\Tick},
3/{\sum_{c,d}(b\soplus c, a)(a, c)(d, 0)}/{\Swap, \Swap},
4/{\sum_{c,d}(b\soplus c, 0)(a,b\soplus d)(d,a)}/{\Tick, \Tick},
5/{\sum_{c,d}(b \soplus c, 0)(b \soplus d, a) (a, d)}/{\Swap, \Swap},
6/{\sum_{c,d}(b \soplus c, 0)(b \soplus d, 0) (a, b)}/{\Tick}}
{
\node [programlabel] (1) at (0, \y) {$\l$};
\draw [hor] (1.east -| 14,0) to (1.east);
\node [programlabel] at (0, \y-0.5) {\t};
}
\end{scope}
\node at (0.5,-0.5) {Alice};
\node at (3,-0.5) {Bob};
\node at (5.5,-0.5) {Charlie};
\end{tz}
\quad=\quad
\def\bottom{-2}
\begin{tz}[protocoldiagram,xscale=0.9]
\path [background] (-0.5,\bottom-1) rectangle +(2,8.5);
\path [background] (3.5,\bottom-1) rectangle +(2,8.5);
\path[surface,even odd rule] (1,\bottom) to [out=up, in=down, out looseness=0.9, in looseness=1.1] (5,5) to (4,5) to [out=down, in=up, out looseness=0.9, in looseness=1.1] (0,\bottom);
\draw[edge] (1,\bottom) to [out=up, in=down, out looseness=0.9, in looseness=1.1] (5,5) (4,5) to [out=down, in=up, out looseness=0.9, in looseness=1.1] (0,\bottom);
\path[surface, draw,edge] (0,5) to [out=down, in=left] (0.5, 4) to [out=right, in=down] (1, 5);
\path[surface, draw,edge] (2,5) to [out=down, in=left] (2.5, 4) to [out=right, in=down] (3, 5);
\node at (0.5,-2.5) {Alice};
\node at (2.5,-2.5) {Bob};
\node at (4.5,-2.5) {Charlie};
\end{tz}
\]

\figurecaptionsuck
\caption{State transfer.}
\label{fig:statetransfer}
\figurecaptionpostsuck
\end{figure*}

Every groudit has its associated dit, with the logical states of the groudit corresponding to the elements of the dit. They interact via the 2\-morphisms depicted in \autoref{fig:measurement}(a) and (b). These are not physical elements of the groudit programming language (explaining why they do not appear in Section~\ref{sec:grouditsanddits}), but auxiliary mathematical structures that we will use to verify our groudit programs.  In \autoref{fig:measurement}(a) we begin with a groudit, and we read it to extract some classical data indicated by the yellow region; the groudit itself still exists. %

Semantically, the blue region represents a groudit $\mathcal G$ with underlying groupoid \G, and the yellow region represents a classical dit $\mathcal B$ with underlying groupoid \cat B, such that \cat B is a discrete groupoid with the same set of objects as \G. We define the yellow-blue morphism $S : \cat B \hto \cat G$ as follows, where the set $\emptyset$ is the empty set, and where $S_{b,g}$ are equipped with the `empty action' and the right action of $\Aut_\cat G(g)$ on itself, respectively:
\begin{equation}
\label{eq:measurementsemantics1}
S_{b,g} :=
\begin{cases}
\Aut_\G(g) & \text{if $b=g$}
\\
\emptyset & \text{otherwise}
\end{cases}
\end{equation}
The blue-yellow morphism $S^* : \cat G \hto \cat B$ is defined similarly. We define \autoref{fig:measurement}(a) as follows, for all $a \in \Ob(\G)=\Ob(\cat B)$ and $g \in \Aut_\G(a)$:
\begin{align}
\label{eq:isospan}
&\text{\autoref{fig:measurement}(a)}
&
g &\mapsto (a, g)
\end{align}
\ignore{
\\
\label{eq:measurementsemantics3}
&\text{\autoref{fig:measurement}(b)}
&
\id_a &\mapsto \textstyle\sum_{g \in \Aut_\G(a)} (\id_a, g)
\end{align}}
The span~\eqref{eq:isospan} is unitary, and hence satisfies equations \autoref{fig:measurement}(d) and (f); equation \autoref{fig:measurement}(e) can be verified analogously. By way of warning, \autoref{fig:measurement}(c) shows a \textit{nonequation} that is not satisfied in general.
\begin{figure}[b!]
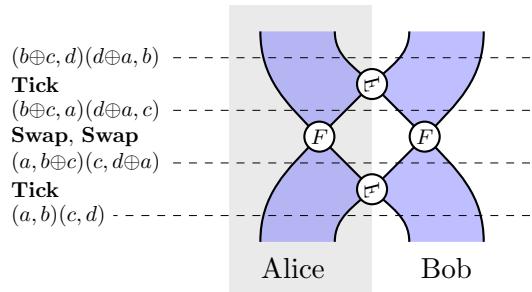

\figuretopsuck
\begin{center}
$\begin{tz}[protocoldiagram]
\path [background] (-\hrt,0) rectangle +(2+\hrt,5.5);
\begin{scope}[scale=2]
\path[surface,even odd rule]
    (1-0.5*\hrt,0.5) to [out=up, in=-135] (1,1) node[blob,rotate=-90,yscale=-1] {$F$} to (1.5,1.5) to [out=45, in=down] (1+1.5*\hrt,2.5) to (1+0.5*\hrt,2.5) to [out=down, in=45] (1,2) to (0.5,1.5) node[blob] {$F$} to [out=-135, in=up] (1-1.5*\hrt,0.5)
    (1-1.5*\hrt,2.5) to [out=down, in=135]  (0.5,1.5) to (1,1) to [out=-45, in=up] (1+0.5*\hrt,0.5) to (1+1.5*\hrt,0.5) to [out=up, in=-45] (1.5,1.5) node [blob] {$F$} to (1,2) node [blob,rotate=-90,yscale=-1] {$F$} to [out=135, in=down] (1-0.5*\hrt,2.5)
    ;
\path[edge]
    (1-0.5*\hrt,0.5) to [out=up, in=-135] (1,1) to (1.5,1.5) to [out=45, in=down] (1+1.5*\hrt,2.5)(1+0.5*\hrt,2.5) to [out=down, in=45] (1,2) to (0.5,1.5) to [out=-135, in=up] (1-1.5*\hrt,0.5)
    (1-1.5*\hrt,2.5) to [out=down, in=135]  (0.5,1.5) to (1,1) to [out=-45, in=up] (1+0.5*\hrt,0.5) (1+1.5*\hrt,0.5) to [out=up, in=-45] (1.5,1.5) to (1,2) to [out=135, in=down] (1-0.5*\hrt,2.5)
    ;
\end{scope}
\begin{scope}[xshift=-5cm, yshift=1.5cm]
\foreach \y/\l/\t in
{0/{(a,b)(c,d)}/{},
1/{(a,b \soplus c) (c,d \soplus a)}/{\Tick},
2/{(b \soplus c, a)(d \soplus a, c)}/{\Swap, \Swap},
3/{(b \soplus c, d)(d \soplus a, b)}/{\Tick}}
{
\node [programlabel] (1) at (0, \y) {$\l$};
\draw [hor] (1.east -| 10,0) to (1.east);
\node [programlabel] at (0, \y-0.5) {\t};
}
\end{scope}
\node at (0.5,0.5) {Alice};
\node at (2+2*\hrt,0.5) {Bob};
\end{tz}$
\end{center}

\figurecaptionsuck
\caption{The basic state transfer repeating block.}
\label{fig:basicstatetransfer}
\figurecaptionpostsuck
\end{figure}


Given the topological behaviour encoded in \autoref{fig:measurement}(b), we can be relaxed about how we draw the interface between yellow and blue regions:
{
\tikzset{every picture/.style={yscale=0.6}}
\begin{equation}
\begin{tz}
\draw [classical] (0,0) rectangle +(1,1.25);
\draw [surface] (0,1.25) rectangle +(1,1.25);
\draw [edge] (0,0) to (0,2.5);
\draw [edge] (1,0) to +(0,2.5);
\draw [edge] (0,1.25) to +(1,0);
\end{tz}
:=
\begin{tz}[xscale=0.666, yscale=1.25]
\draw [classical, edge] (0,0) to (0,0.5) to [out=up, in=-135] (0.5,1.5) to [out=-45, in=135] (1.5,0.5) to (1.5,0);
\draw [surface, edge] (0.5,2) to (0.5,1.5) to (1.5,0.5) to [out=45, in=down] (2,1.5) to (2,2);
\end{tz}
=
\begin{tz}[xscale=0.666, yscale=-1.25]
\draw [surface, edge] (0,0) to (0,0.5) to [out=up, in=-135] (0.5,1.5) to [out=-45, in=135] (1.5,0.5) to (1.5,0);
\draw [classical, edge] (0.5,2) to (0.5,1.5) to (1.5,0.5) to [out=45, in=down] (2,1.5) to (2,2);
\end{tz}
\end{equation}}%
This gives us our composite \Write operation; \Read is the dagger of this. We also use this to define yellow-blue and yellow-yellow versions of the biunitary in \autoref{fig:redredbiunitary}(a)--(c). 
\begin{proposition} Equation~\autoref{fig:redredbiunitary}(d) is fulfilled by the biunitary associated to any groudit.
\end{proposition}
\begin{proof} The composite $2$-morphism on the left hand side of \autoref{fig:redredbiunitary}(d) maps a dit $[a]$ to the multiset $\sum_{x\in \Aut_{\cat{G}}(a)} [s(F_{\epsilon,\tau}(x))]$, where $F_{\epsilon, \tau}$ is the function~\eqref{eq:associatedbiunitary} and $s$ denotes the source of the morphism $F_{\epsilon,\tau}(x)$. It follows from the explicit expression~\eqref{eq:associatedbiunitary} that $s(F_{\epsilon,\tau}(x)) = \epsilon_{a}(x)$. Since $\epsilon_a: \Aut_{\cat{G}}(a) \to \Ob(\cat{G})$ is a bijection, $[a]$ is mapped to the multiset $\sum_{x\in \Aut_{\cat{G}}(a)} [\epsilon_{a}(x)] = \sum_{b \in \Ob(\cat{G})} [b]$ which is the right hand side of~\autoref{fig:redredbiunitary}(d).
\end{proof}
\noindent
Equation~\autoref{fig:redredbiunitary}(e) is a direct consequence of horizontal unitarity of $F$. These structures yield our composite operations \IRead, \IWrite, \LRead, \RRead and \CTick.

\section{Protocols}
\label{sec:protocols}

\subsection{State transfer (Figures \ref{fig:statetransfer} and \ref{fig:basicstatetransfer})}
\label{sec:statetransfer}

\firstparagraph{Overview.} The state transfer protocol communicates a groubit down a linear chain of nodes, such that each node is connected to its neighbour with a link. Our mathematical treatment is closely related to a state transfer protocol for cluster-based quantum computers proposed previously by the authors~\cite{Reutter:2017a}. The adjective \textit{timeless} arises from a specific property of this protocol, which we examine below.

\paragraph{Program.} The state transfer program is illustrated in \autoref{fig:statetransfer}(b) for three parties, Alice, Bob and Charlie, arranged in a linear chain. Each party has a node, and separate links connect Alice and Bob, and also Bob and Charlie, enabling \Tick operations between connected parties. Alice has a groubit, which she would like to transfer to Charlie coherently; that is, preserving the internal state. The protocol is formed from repetitions of the \textit{basic scheme} (see \autoref{fig:basicstatetransfer}), involving a \Tick operation, two \Swap operations, and a final \Tick.
In \autoref{fig:statetransfer}(a) we use 2 copies of this basic building block, one between Alice and Bob, and one between Bob and Charlie. The generalization to arbitrary linear chains is clear.

\paragraph{Verification.}
The protocol is verified in the general case by observing that \autoref{fig:statetransfer}(a) can be transformed into \autoref{fig:statetransfer}(b) by applying the equations of \autoref{fig:biunitarity}. On the left-hand side of \autoref{fig:statetransfer}(a) we give an explicit program trace for the case of a $\Z_2\sqcup\Z_2$ groubit, based on the lookup table in Section~\ref{sec:grouditsanddits}. The final state is
$\textstyle\sum_{c,d \in {\Z_2}}(b \soplus c, 0)(b \soplus d, 0) (a, b)$;
by a simple change of variables it is clear that this equals
$\textstyle\sum_{c,d \in \Z_2}(c,0)(d,0)(a,b)$
as required.

\paragraph{Discussion.}
This protocol has certain limitations. While multiple messages can be sent from left-to-right along such a linear chain of nodes, if one attempts to send a message from right-to-left at the same time using a reflected version of the protocol, then both messages will be corrupted. Of course, this could be overcome by having a pair of parallel chains, keeping left-to-right and right-to-left communications on separate tracks. Furthermore, we do not have a clear analysis of communication on a network with a more interesting topology.

\paragraph{Timelessness.}
A key property of this protocol is that it makes no use of timeouts, thanks to atomicity properties that are part of our basic assumptions (see Section~\ref{sec:overview}.) This is desirable, since timeouts are a basic feature of the dominant TCP protocol for internet communication~\cite{Iren_1999} which are the source of reliability issues in data centre environments~\cite{Borrill:2016, Adesanmi_2015}. If the final \Tick event of the scheme given in \autoref{fig:basicstatetransfer} succeeds, then Bob assumes ownership of the message and continues to propagate it. Otherwise, if the final \Tick is not successful---which could be because one of the 3 earlier events were not successful---Alice maintains ownership of the message, and is free to direct it by another route, or to return it as undeliverable to the sender.

\begin{figure}[b!]
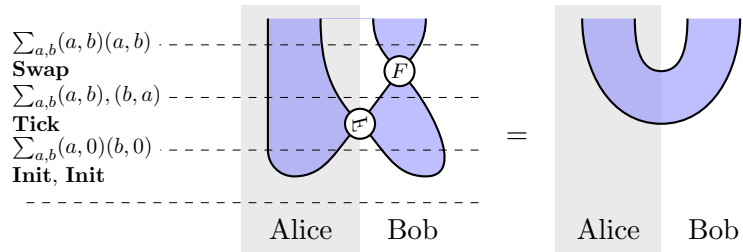

\figuretopsuck
\begin{center}
$\begin{tz}[protocoldiagram]
\path[background] (-1.5,-2.5) rectangle +(2.25,4.75);
\node at (-0.375,-2) {Alice};
\node at (1.75,-2) {Bob};

\path [surface, even odd rule, postaction={draw, edge}] (-1, 2) to [out=down, in=up] (-1,-0.5) to [out=down, in=left] (-0.5,-1) to [out=right, in=-135] (0.75,0) node (F1) [blob,rotate=-90,yscale=-1] {$F$} to (1.5,1) node (F2) [blob] {$F$} to [out=45, in=down] (2,2)
 (0,2) node (4) {} to [out=down, in=135] (F1.center) to [out=-45, in=left] (2,-1) to [out=right, in=-45] (F2.center) to [out=135, in=down] (1,2);

\def\leftmost{-6cm}
\foreach \y/\l/\t in
{-1.5/{}/{},
-0.5/{\sum_{a,b}(a,0)(b,0)}/{\Init, \Init},
0.5/{\sum_{a,b}(a,b),(b,a)}/{\Tick},
1.5/{\sum_{a,b}(a,b)(a,b)}/{\Swap}}
{
\node [programlabel] (1) at (\leftmost, \y) {$\l$};
\draw [hor] (1.east -| 3,0) to (1.east);
\node [programlabel] at (\leftmost+0.0cm, \y-0.5) {\t};
}

\end{tz}
\,\,\,=\,\,\,
\begin{tz}[protocoldiagram]  
\path[background] (-2,-2.5) rectangle +(2,4.75);
\node at (-1,-2) {Alice};
\node at (1,-2) {Bob};

\path[surface, even odd rule, postaction={edge}] 
(-1.5, 2) to [out=down, in=left]  (0,0) to [out=right, in=down] (1.5,2)
    (0.5,2) to [out=down, in=right]  (0, 1) to [out=left, in=down] (-0.5,2);
 
\end{tz}$
\end{center}

\figurecaptionsuck
\caption{Entanglement creation.}
\label{fig:entanglementcreation}
\figurecaptionpostsuck
\end{figure}

\subsection{Entanglement creation (\autoref{fig:entanglementcreation})}
\label{sec:entanglementcreation}

\firstparagraph{Overview.}
This is a procedure to create an `entangled pair' of groubits. Entangled groubits are required for the dense coding and teleportation protocols described later.

\paragraph{Program.}
Alice and Bob each initialize a groubit. They then perform a \Tick operation involving both their groubits. Finally, Bob performs a \Swap operation.

\paragraph{Verification.}
Immediate by \autoref{fig:biunitarity}(a).

\paragraph{Discussion.}
To implement this protocol, Alice and Bob must be connected by a link enabling the \Tick operation.

\begin{figure*}
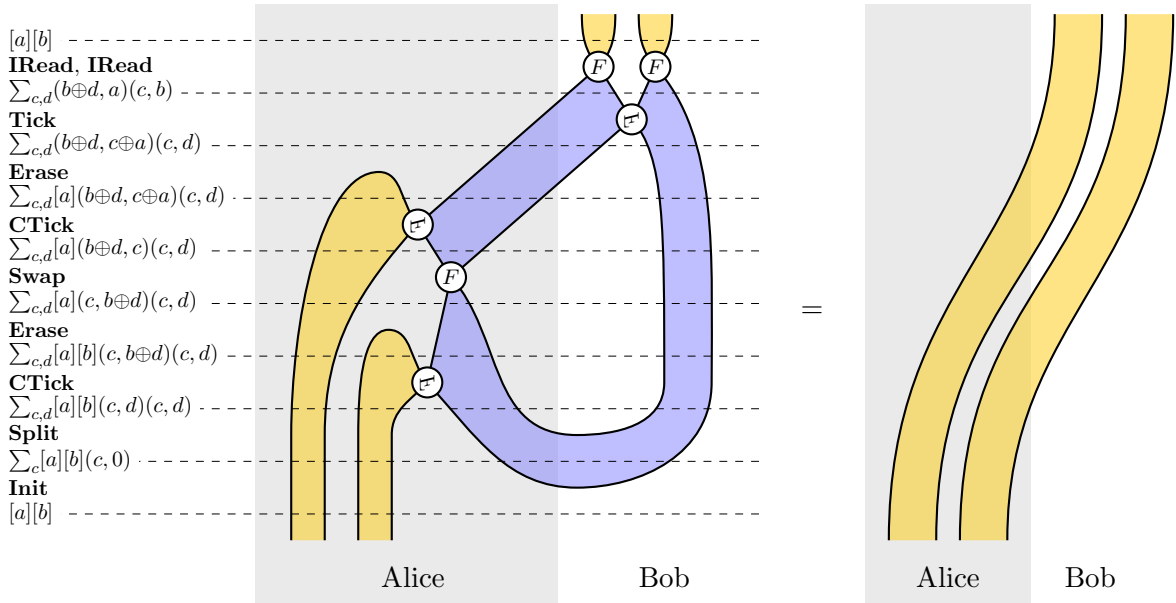

\begin{center}
\mbox{
\hspace{-4.5cm}
$\begin{tz}[protocoldiagram,xscale=0.9]
\clip(-8,-3.2) rectangle(8,8.2);
\node[blob,rotate=-90, yscale=-1](Y) at (1,1) {$F$};
\node[blob] (F1) at (1.5,3) {$F$};
\node[blob,rotate=-90, yscale=-1](X) at (0.8,4) {$F$};

\node[blob](C) at  (4.4+2*\hrt, 7){$F$};
\node[blob,rotate=-90, yscale=-1](B) at  (3.9+2*\hrt, 6) {$F$};
\node[blob](A) at  (3.2+2*\hrt, 7){$F$};
\def \top {8}
\node (T4) at (4.4+2.5*\hrt, \top){};
\node (T3) at ([xshift=-\hrt cm] T4){};
\node (T2) at (3.2+2.5*\hrt, \top){};
\node (T1) at ([xshift=-\hrt cm] T2){};
\def \w{1.7}
\def \d{1.5}
\def \middle {1.75+\hrt+\w}
\node (M) at (\middle, -1){};
\node (M2) at ([yshift=1cm] M.center){};
\node (R) at (6, 1){};
\node (R2) at (7,1){};
\def\bottom{-2}

\path[background] (-0.5-3*\hrt, \bottom-1.2) rectangle (\middle-0.4, \top + 0.2);
\node at (0.7, \bottom-0.7) {$\mathrm{Alice}$};
\node at (6., \bottom-0.7) {$\mathrm{Bob}$};

\path[surface, edge,draw] (F1.center) to (B.center) to (A.center) to (X.center) to (F1.center);
\path[surface,edge,draw] (F1.center) to (Y.center)
    to [out=-45, in=left] (M.center) to [out=right, in=down] (R2.center) to [out=up, in=-45, in looseness=0.8] (C.center) to (B.center) to [out=-45, in=up, out looseness=0.5, in looseness=1.5] (R.center) to [out=down, in=right] (M2.center) to [out=left, in=-45, out looseness=1.2] (F1.center);
\path[classical,edge]  (0.25-\hrt,-2)to (0.25-\hrt, 0) to [out=up, in=left, in looseness=0.7] (0.25-0.1*\hrt,2) to [out=right,in=135,out looseness=1] (1,1)  to [out=-135, in=up] (0.25,0) to (0.25,\bottom);
\path[classical, edge] (0.25-3*\hrt, -2) to  (0.25-3*\hrt, 0) to [out=up, in=left, in looseness=0.6] (0.05-0.1*\hrt, 5) to [out=right, in=135, out looseness=1](X.center) to [out=-135, in=up] (0.25-2*\hrt,0) to  (0.25-2*\hrt,-2);
\path[classical, edge] (T1.center) to [out=down, in=135] (A.center) to [out=45, in=-90] (T2.center);
\path[classical,edge] (T3.center) to [out=down, in=135] (C.center) to [out=45, in=down] (T4.center);

\path[partition] (\middle,\bottom-0.5) to (\middle, \top +0.5);

\def\leftmost{-8cm}
\foreach \y/\l/\t in
{-1.5/{[a][b]}/{},
-0.5/{\sum_c[a][b] (c,0)}/{\Init},
0.5/{\sum_{c,d}[a][b](c,d) (c,d)}/{\Split},
1.5/{\sum_{c,d}[a][b](c,b\soplus d) (c,d)}/{\CTick},
2.5/{\sum_{c,d}[a](c,b\soplus d) (c,d)}/{\Erase},
3.5/{\sum_{c,d}[a](b\soplus d, c) (c,d)}/{\Swap},
4.5/{\sum_{c,d}[a](b\soplus d, c\soplus a) (c,d)}/{\CTick},
5.5/{\sum_{c,d}(b\soplus d, c\soplus a) (c,d)}/{\Erase},
6.5/{\sum_{c,d}(b\soplus d, a) (c,b)}/{\Tick},
7.5/{[a][b]}/{\IRead, \IRead}}
{
\node [programlabel] (1) at (\leftmost, \y) {$\l$};
\draw [hor] (1.east -| 8,0) to (1.east);
\node [programlabel] at (\leftmost+0.0cm, \y-0.5) {\t};
}
\end{tz}
\quad=\quad
\begin{tz}[protocoldiagram,xscale=0.9]
\def \top {5}
\node (T4) at (3.4+2.5*\hrt, \top){};
\node (T3) at ([xshift=-\hrt cm] T4){};
\node (T2) at (2.2+2.5*\hrt, \top){};
\node (T1) at ([xshift=-\hrt cm] T2){};
\def \d{1.5}
\def\bottom{-\top}

\path[background] (-2, \bottom-1.2) rectangle (1.5, \top + 0.2);

\node at (-0.25, \bottom-0.7) {$\mathrm{Alice}$};
\node at (2.75, \bottom-0.7) {$\mathrm{Bob}$};
\def\aloose{1.2}
\def\bloose{1.2}
\def\cloose{1.2}
\def\dloose{1.05}
\def\xgap{3.5}

\path[classical] (0,\bottom) to [out=up, in=down, out looseness=\aloose, in looseness=\bloose] (\xgap,\top) to (\xgap+1,\top) to [out=down, in=up, out looseness=\cloose, in looseness=\dloose] (1,\bottom);
\path[edge] (0,\bottom) to [out=up, in=down, out looseness=\aloose, in looseness=\bloose] (\xgap,\top);
\draw[edge] (\xgap+1,\top) to [out=down, in=up, out looseness=\cloose, in looseness=\dloose] (1,\bottom);

\begin{scope}[xshift=3cm, yshift=0cm, rotate=180]
\path[classical] (0,\bottom) to [out=up, in=down, out looseness=\aloose, in looseness=\bloose] (\xgap,\top) to (\xgap+1,\top) to [out=down, in=up, out looseness=\cloose, in looseness=\dloose] (1,\bottom);
\path[edge] (0,\bottom) to [out=up, in=down, out looseness=\aloose, in looseness=\bloose] (\xgap,\top);
\draw[edge] (\xgap+1,\top) to [out=down, in=up, out looseness=\cloose, in looseness=\dloose] (1,\bottom);
\end{scope}
 

\end{tz}$
\hspace{-5cm}}
\end{center}

\figurecaptionsuck
\caption{Dense coding.}
\label{fig:densecoding}
\figurecaptionpostsuck
\end{figure*}

\subsection{Dense coding (\autoref{fig:densecoding})}
\label{sec:densecoding}

\firstparagraph{Overview.} The dense coding procedure allows 2 classical bits to be transmitted between two parties, by transferring only 1 groubit. The parties must share an entangled pair of groubits, which could have been generated by the  procedure discussed in Section~\ref{sec:entanglementcreation}.

\paragraph{Program.} Alice begins with two classical bits, and Alice and Bob share an entangled pair of groubits. Alice begins by performing \CTick operations (see Section~\ref{sec:grouditsanddits}) between her classical bits and her groubit, with a \Swap operation in between. She then transfers the groubit to Bob, who performs a \Tick operation between his two groubits, and then \IRead operations on both groubits.

\paragraph{Verification.} To verify correctness of the program for general groudits, substitute the definitions of \IRead and \CTick in terms of the basic syntax, then apply equations from \autoref{fig:biunitarity} to cancel 3 pairs of adjacent $F$ nodes.

\paragraph{Discussion.}
It may seem surprising that dense coding is possible, since although a groubit has 2 classical bits of memory, they cannot both be directly accessed; applying the \Read operation (see Section~\ref{sec:grouditsanddits}) reveals the logical bit, but destroys the internal bit. The program requires passing a groubit from one agent to another; to implement this, agents could use the state transfer program described in Section~\ref{sec:statetransfer}.

Dense coding allows agents connected by a groubit network to double their effective data transfer rate, at the expense of consuming shared entanglement. It may be possible to use this for temporal load-balancing in a groubit data center. During times of low utilization, agents in the network perform entanglement creation (Section~\ref{sec:entanglementcreation}) to generate substantial numbers of shared entangled groubits. Later, when utilization of the data centre becomes high, these entangled groubits can be consumed to double the effective rate of data transfer.

\begin{figure*}
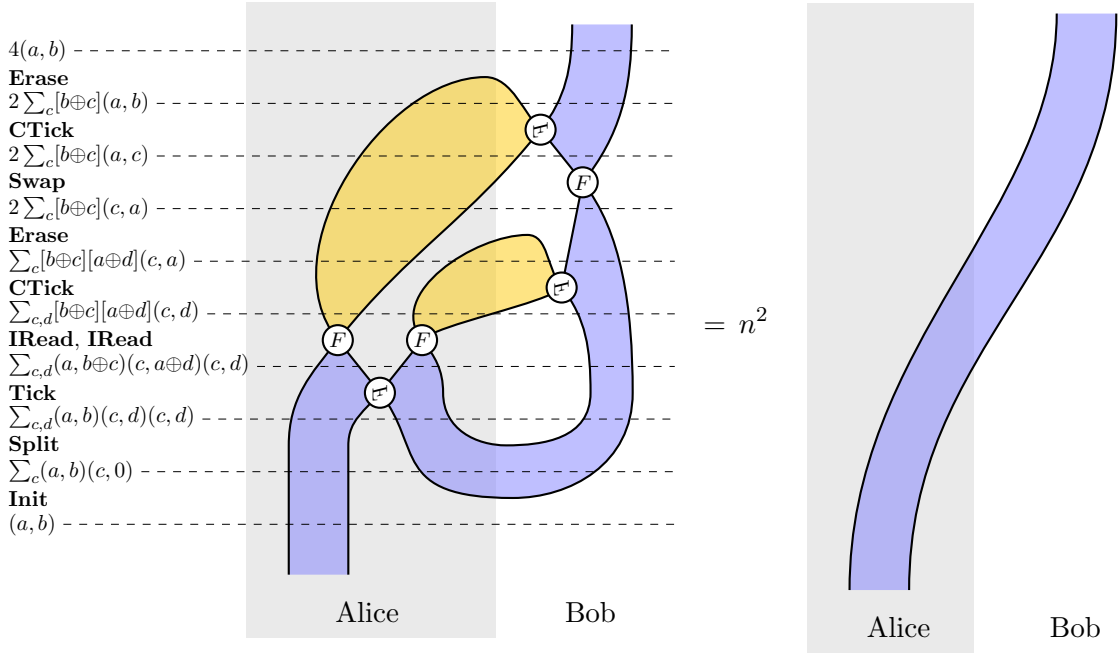

\figuretopsuck
\[ \hspace{-20pt}
\begin{tz}[protocoldiagram,xscale=0.8]
\node[blob,rotate=-90, yscale=-1] (A) at (1,1) {$F$};
\node[blob](B) at (2,2) {$F$};
\node[blob](E) at (0.,2) {$F$};
\node[blob](D) at  (4.4+2*\hrt, 5){$F$};
\node[blob,rotate=-90, yscale=-1](C) at  (3.9+2*\hrt, 3) {$F$};
\node[blob,rotate=-90, yscale=-1](F) at  (3.4+2*\hrt, 6){$F$};
\def \top {4+2*\hrt}
\node (Q) at (4.5,4){};
\node (S) at (3.5, 7){};
\def \leftb{-0.75-2*\hrt}
\def\rightb{4.15+4*\hrt}

\node (T1) at (4.15+2*\hrt, 8){};
\node (T2) at ([xshift=1.414 cm] T1.center){};

\def \w{1.7}
\def \d{1.5}
\def \middle {1.75+\hrt+\w}
\node (M) at (\middle, -1){};
\node (M2) at ([yshift=\hrt cm] M.center){};
\node (R) at (1.75+\hrt+2*\w, 0){};
\node (R2) at (7,1){};
\node (TR) at (2.65+2*\hrt, \top) {};
\node (TL) at (2.65-2*\w, \top){};
\def\bottom{-\d-0.25-\hrt}

\path[background] (-0.75-2*\hrt, \bottom-1.2) rectangle (\middle-0.4, \top + 3);
\node at (0.7, \bottom-0.7) {$\mathrm{Alice}$};
\node at (6., \bottom-0.7) {$\mathrm{Bob}$};

\path[surface,edge] (A.center) to [out=-45, in=left, in looseness=2] (M.center) to [out=right, in=down, out looseness=1] (R2.center) to [out=up, in=-45, in looseness=0.8] (D.center) to (C.center) to [out=-45, in=up]  (6,1) to [out=down, in=right] (4,0) to [out=left, in=down] (2.5,1) to [out=up, in=-45] (B.center) to (A.center);

\path[surface, edge] (0.25-2*\hrt,\bottom) to  (0.25-2*\hrt,0) to [out=up, in=-135](E.center)  to (A.center) to [out=-135, in=up] (0.25,0) to (0.25,\bottom);

\path[surface,edge] (T1.center) to [in=45, out=down] (F.center) to (D.center) to  [out=45, in=down] (T2.center);

\path[classical, edge]  (C.center) to [out=135, in=right, out looseness=1, in looseness=1] (Q.center) to [out=left, in=135, in looseness=1]  (B.center)  to[out=45, in=-135,looseness=0.3] (C.center);

\path[classical,edge] (F.center) to [out=145, in=right, out looseness=0.7, in looseness=1] (S.center) to [out=left, in=135, in looseness=1]  (E.center) to  [out=45, in=-135] (F.center);

\path[partition] (\middle-0.3,\bottom-0.5) to (\middle-0.3, \top +0.5);

\def\leftmost{-8cm}
\foreach \y/\l/\t in
{-1.5/{(a,b)}/{},
-0.5/{\sum_c(a,b) (c,0)}/{\Init},
0.5/{\sum_{c,d}(a,b) (c,d) (c,d)}/{\Split},
1.5/{\sum_{c,d}(a,b\soplus c) (c,a\soplus d) (c,d)}/{\Tick},
2.5/{\sum_{c,d}[b\soplus c] [a\soplus d] (c,d)}/{\IRead, \IRead},
3.5/{\sum_{c}[b\soplus c][a \soplus d] (c,a)}/{\CTick},
4.5/{2\sum_{c}[b\soplus c] (c,a)}/{\Erase},
5.5/{2\sum_{c}[b\soplus c] (a,c)}/{\Swap},
6.5/{2\sum_{c}[b\soplus c] (a,b)}/{\CTick},
7.5/{4 (a,b)}/{\Erase}}
{
\node [programlabel] (1) at (\leftmost, \y) {$\l$};
\draw [hor] (1.east) to (1.east -| 8,0);
\node [programlabel] at (\leftmost+0.0cm, \y-0.5) {\t};
}
\end{tz}
\hspace{-5pt}\quad=\,n^2\,\quad
\begin{tz}[protocoldiagram,xscale=0.8]
\def \top {8.5}
\node (T4) at (3.4+2.5*\hrt, \top){};
\node (T3) at ([xshift=-2*\hrt cm] T4){};
\def \d{1.5}
\def\bottom{-\d-0.25-\hrt}

\path[background] 
(-0.75-2*\hrt, \bottom-1.2) rectangle (1.25+0.75*\hrt,\top+0.2);

\node at (0., \bottom-0.7) {$\mathrm{Alice}$};
\node at (4.2, \bottom-0.7) {$\mathrm{Bob}$};
\path[surface]
 (0.25-2*\hrt,\bottom) to [out=up, in=down, out looseness=1, in looseness=0.8] (T3.center) to (T4.center) to [out=down, in=up] (0.25,\bottom);
 
 \ignore{
\path[classical,edge] (0.15-1.5*\hrt, \top ) to [out=down, in=left]  (1.325-0.5*\hrt,\top-1.5) to [out=right, in=down] (2.5+0.5*\hrt,\top);
 }
 
 \draw[edge]  (0.25-2*\hrt,\bottom) to [out=up, in=down, out looseness=1, in looseness=0.8] (T3.center) ;

 \draw[edge]  (T4.center) to [out=down, in=up] (0.25,\bottom);
\path[partition] (1.25+0.75*\hrt,\bottom-0.5) to (1.25+0.75*\hrt,\top +0.5);

\end{tz}
\]

\figurecaptionsuck
\caption{Teleportation.}
\label{fig:teleportation}
\figurecaptionpostsuck
\end{figure*}

\subsection{Teleportation (\autoref{fig:teleportation})}
\label{sec:teleportation}

\firstparagraph{Overview.}
The teleportation procedure allows a groubit to be transported from one location to another, as long as those locations share an entangled groubit pair (see Section~\ref{sec:entanglementcreation}.) 

\paragraph{Program.}
There are two parties, Alice and Bob. Alice starts with a groubit to be teleported, and Alice and Bob share between them an entangled pair of groubits. First, Alice performs a \Tick operation on the groubit to be teleported. She then performs \Swap operations on both of her groubits, then converts them into classical bits, which are transmitted to Bob by conventional means (for example, over the internet.) Bob then performs two \CTick operations (see \autoref{fig:measurement}), and performs \Erase on the classical data received from Alice. The result is that Bob's groubit is now in the same state as Alice's was originally, both with respect to its logical and internal data.

\paragraph{Verification.}
To verify the protocol in the general case, expand the \CTick operations using the definitions from \autoref{fig:redredbiunitary}(a) and (b), then apply equation~\autoref{fig:redredbiunitary}(e) twice. The result is the identity, up to two yellow bubbles, which count the different classical bits that Alice could have obtained.

\paragraph{Discussion.}
Teleportation may have an application for transferring groubits between separate groubit networks, which may only be connected via the internet. Of course, these data centres would have to be furnished with a sufficient supply of entangled groudits.

\begin{figure*}[b]
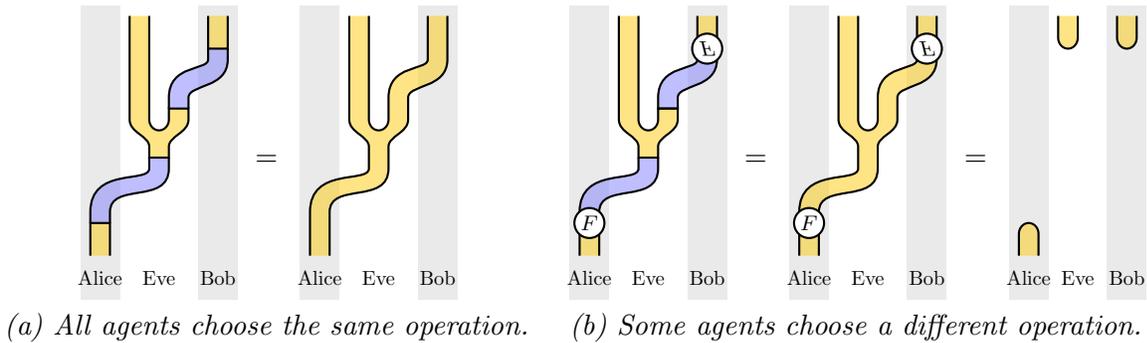

\def\scl{0.29}
\def\yscl{1.0}
\def\xscl{0.9}
\def\nameheight{-2}
\def\sclblob{1}
\def\sclnames{0.7}
\[
\hspace{-0.95cm}
\vc{$\begin{tz}[scale=\scl,yscale=\yscl,xscale=\xscl]
\path[background] (-1.5,-3) rectangle  (0.5,10.5);
\path[background] (4.5,-3) rectangle  (6.5,10.5);
\node [scale=\sclnames] at (-0.5,\nameheight) {$\mathrm{Alice}$};
\node [scale=\sclnames] at (2.5,\nameheight) {$\mathrm{Eve}$};
\node [scale=\sclnames] at (5.5,\nameheight){$\mathrm{Bob}$};
\path[surface] (5,8.5) to (5,8) to [out=down, in=up, in looseness=1.5,out looseness=0.5] (3,6.25) to (3,5.75) to (4,5.75) to   (4,6.25) to [out=up, in=down, out looseness=0.6, in looseness=1.3] (6,8) to (6,8.5);
\draw[edge](5,8.5) to (5,8) to [out=down, in=up, in looseness=1.5,out looseness=0.5] (3,6.25) to (3,5.75);
\draw[edge] (4,5.75) to  (4,6.25) to [out=up, in=down, out looseness=0.6, in looseness=1.3] (6,8) to (6,8.5);
\path[surface](-1,0.5) to (-1,1) to [out=up, in=down,out looseness=1.5, in looseness=0.5] (2, 3) to(2,3.5) to (3,3.5) to (3,3)to [out=down, in=up, out looseness=1.2] (0,1) to (0,0.5);
\draw[edge] (-1,0.5) to (-1,1) to [out=up, in=down,out looseness=1.5, in looseness=0.5] (2, 3) to(2,3.5);
\draw[edge](3,3.5) to (3,3)to [out=down, in=up, out looseness=1.2] (0,1) to (0,0.5);
\path[classical,draw,edge] (-1,-1) to (-1,0.5) to (0,0.5) to (0,-1);
\path[classical,draw,edge] (1,10) to [out=down, in=up] (1,5.25) to [out=down, in=up, in looseness=0.8] (2,4) to (2,3.5) to (3,3.5)to (3,4)to  [out=up, in=down, out looseness=0.8] (4,5.25) to (4,5.75) to (3,5.75) to (3,5.25) to [out=down, in=right] (2.5,4.75)  to [out=left, in=down] (2,5.25) to [out=up, in=down] (2,10);
\path[classical, draw,edge] (5,10) to (5,8.5) to (6,8.5) to (6,10);
\path[partition] (0.5,-1.5) to (0.5,10.5);
\path[partition](4.5,-1.5) to  (4.5,10.5);
\end{tz}
=
\begin{tz}[scale=\scl,yscale=\yscl,xscale=\xscl]
\path[background] (-1.5,-3) rectangle  (0.5,10.5);
\path[background] (4.5,-3) rectangle  (6.5,10.5);
\node [scale=\sclnames] at (-0.5,\nameheight) {$\mathrm{Alice}$};
\node [scale=\sclnames] at (2.5,\nameheight) {$\mathrm{Eve}$};
\node [scale=\sclnames] at (5.5,\nameheight){$\mathrm{Bob}$};
\path[classical]
(-1,-1) to  (-1,1) to [out=up, in=down,out looseness=1.5, in looseness=0.5] (2, 3) to (2,4) to [out=up, in=down, out looseness=0.8] (1,5.25)  to [out=up, in=down] (1,10) to (2, 10) to [out=down, in=up]  (2,5.25) to [out=down, in=left] (2.5,4.75) to [out=right, in=down] (3,5.25) to (3,6.25) to [out=up, in=down, out looseness=1.5, in looseness=0.5] (5,8.) to (5,10) to (6,10) to (6,8) to [out=down, in=up, out looseness=1.3, in looseness=0.6] (4,6.25) to (4,5.25) to [out=down, in=up, in looseness=0.8] (3,4) to (3,3) to [out=down, in=up, out looseness=1.2] (0,1) to (0,-1);
\draw[edge] (-1,-1) to  (-1,1) to [out=up, in=down,out looseness=1.5, in looseness=0.5] (2, 3) to (2,4) to [out=up, in=down, out looseness=0.8] (1,5.25)  to [out=up, in=down] (1,10) ;
\draw[edge] (0.,-1) to (0.,1) to [out=up, in=down, in looseness=1.2] (3,3) to (3,4) to  [out=up, in=down, out looseness=0.8] (4,5.25) to (4,6.25) to [out=up, in=down, out looseness=0.6, in looseness=1.3] (6,8) to (6,10);
\draw[edge] (2, 10) to [out=down, in=up]  (2,5.25) to [out=down, in=left] (2.5,4.75) to [out=right, in=down] (3,5.25) to (3,6.25) to [out=up, in=down, out looseness=1.5, in looseness=0.5] (5,8.) to (5,10);
\path[partition] (0.5,-1.5) to (0.5,10.5);
\path[partition](4.5,-1.5) to  (4.5,10.5);
\end{tz}$
\\
\textit{(a) All agents choose the same operation.}
}
\vc{$\begin{tz}[scale=\scl,yscale=\yscl,xscale=\xscl]
\path[background] (-1.5,-3) rectangle  (0.5,10.5);
\path[background] (4.5,-3) rectangle  (6.5,10.5);
\node [scale=\sclnames] at (-0.5,\nameheight) {$\mathrm{Alice}$};
\node [scale=\sclnames] at (2.5,\nameheight) {$\mathrm{Eve}$};
\node [scale=\sclnames] at (5.5,\nameheight){$\mathrm{Bob}$};
\node[scale=\sclblob,blob] at (-0.5,0.5) {$F$};
\node[scale=\sclblob,blob,yscale=-1] at (5.5,8.5) {$F$};
\path[surface,draw,edge] (5.5,8.5) to [out=-135, in=up] (5,8) to [out=down, in=up, in looseness=1.5,out looseness=0.5] (3,6.25) to (3,5.75) to (4,5.75) to   (4,6.25) to [out=up, in=down, out looseness=0.6, in looseness=1.3] (6,8) to [out=up, in=-45] (5.5,8.5);
\path[surface](-0.5,0.5) to[out=135, in=down]  (-1,1) to [out=up, in=down,out looseness=1.5, in looseness=0.5] (2, 3) to(2,3.5) to (3,3.5) to (3,3)to [out=down, in=up, out looseness=1.2] (0,1) to[out=down, in=45] (-0.5,0.5);
\draw[edge] (-0.5,0.5) to [out=135, in=down]  (-1,1) to [out=up, in=down,out looseness=1.5, in looseness=0.5] (2, 3) to(2,3.5);
\draw[edge](3,3.5) to (3,3)to [out=down, in=up, out looseness=1.2] (0,1) to [out=down, in=45] (-0.5,0.5);
\path[classical,draw,edge] (-1,-1) to (-1,0) to [out=up, in=-135]  (-0.5,0.5) to[out=-45, in=up] (0,0)  to (0,-1);
\path[classical,draw,edge] (1,10) to [out=down, in=up] (1,5.25) to [out=down, in=up, in looseness=0.8] (2,4) to (2,3.5) to (3,3.5)to (3,4)to  [out=up, in=down, out looseness=0.8] (4,5.25) to (4,5.75) to (3,5.75) to (3,5.25) to [out=down, in=right] (2.5,4.75)  to [out=left, in=down] (2,5.25) to [out=up, in=down] (2,10);
\path[classical, draw,edge] (5,10)  to(5,9) to [out=down, in=135] (5.5,8.5)to[out=45, in=down] (6,9) to (6,10);
\path[partition] (0.5,-1.5) to (0.5,10.5);
\path[partition](4.5,-1.5) to  (4.5,10.5);
\end{tz}
=
\begin{tz}[scale=\scl,yscale=\yscl,xscale=\xscl]
\path[background] (-1.5,-3) rectangle  (0.5,10.5);
\path[background] (4.5,-3) rectangle  (6.5,10.5);
\node [scale=\sclnames] at (-0.5,\nameheight) {$\mathrm{Alice}$};
\node [scale=\sclnames] at (2.5,\nameheight) {$\mathrm{Eve}$};
\node [scale=\sclnames] at (5.5,\nameheight){$\mathrm{Bob}$};
\node[scale=\sclblob,blob] at (-0.5,0.5) {$F$};
\node[scale=\sclblob,blob,yscale=-1] at (5.5,8.5) {$F$};
\path[classical,draw,edge] (-1,-1) to (-1,0) to [out=up, in=-135] (-0.5,0.5) to [out=-45, in=up] (0,0) to (0,-1);
\path[classical, draw,edge] (5,10) to (5,9) to [out=down, in=135] (5.5,8.5) to [out=45, in=down] (6,9) to (6,10);
\path[classical]
 (-0.5,0.5) to[out=135, in=down]  (-1,1) to [out=up, in=down,out looseness=1.5, in looseness=0.5] (2, 3) to (2,4) to [out=up, in=down, out looseness=0.8] (1,5.25)  to [out=up, in=down] (1,10) to (2, 10) to [out=down, in=up]  (2,5.25) to [out=down, in=left] (2.5,4.75) to [out=right, in=down] (3,5.25) to (3,6.25) to [out=up, in=down, out looseness=1.5, in looseness=0.5] (5,8.) to [out=up, in=-135] (5.5,8.5) to [out=-45, in=up] (6,8)to [out=down, in=up, out looseness=1.3, in looseness=0.6] (4,6.25) to (4,5.25) to [out=down, in=up, in looseness=0.8] (3,4) to (3,3) to [out=down, in=up, out looseness=1.2] (0,1) to [out=down, in=45] (-0.5,0.5);
\draw[edge] (-0.5,0.5) to [out=135, in=down] (-1,1) to [out=up, in=down,out looseness=1.5, in looseness=0.5] (2, 3) to (2,4) to [out=up, in=down, out looseness=0.8] (1,5.25)  to [out=up, in=down] (1,10) ;
\draw[edge] (-0.5,0.5) to [out=45, in=down] (0.,1) to [out=up, in=down, in looseness=1.2] (3,3) to (3,4) to  [out=up, in=down, out looseness=0.8] (4,5.25) to (4,6.25) to [out=up, in=down, out looseness=0.6, in looseness=1.3] (6,8) to [out=up, in=-45] (5.5,8.5);
\draw[edge] (2, 10) to [out=down, in=up]  (2,5.25) to [out=down, in=left] (2.5,4.75) to [out=right, in=down] (3,5.25) to (3,6.25) to [out=up, in=down, out looseness=1.5, in looseness=0.5] (5,8.) to [out=up, in=-135] (5.5,8.5) ;
\path[partition] (0.5,-1.5) to (0.5,10.5);
\path[partition](4.5,-1.5) to  (4.5,10.5);
\end{tz}
=
\begin{tz}[scale=\scl,yscale=\yscl,xscale=\xscl]
\path[background] (-1.5,-3) rectangle  (0.5,10.5);
\path[background] (3.5,-3) rectangle  (5.5,10.5);
\node [scale=\sclnames] at (-0.5,\nameheight) {$\mathrm{Alice}$};
\node [scale=\sclnames] at (2,\nameheight) {$\mathrm{Eve}$};
\node [scale=\sclnames] at (4.5,\nameheight){$\mathrm{Bob}$};
\draw[classical,draw,edge] (-1,-1) to (-1,0) to [out=up, in=left] (-0.5,0.5) to [out=right, in=up] (0,0) to (0,-1);
\begin{scope}[xshift=-1cm]
\draw[classical,draw,edge] (5,10) to (5,9) to [out=down, in=left] (5.5,8.5) to [out=right, in=down] (6,9) to (6,10);
\end{scope}
\draw[classical,draw,edge] (1,10) to [out=down, in=up] (1,9) to [out=down, in=left] (1.5,8.5) to [out=right, in=down] (2,9) to (2,10);
\end{tz}$
\\
\textit{(b) Some agents choose a different operation.}
}
\]

\figurecaptionsuck
\caption{Verification of BB84 quantum key distribution.}
\label{fig:qkddetailed}
\figurecaptionpostsuck
\end{figure*}

\begin{figure}[b]
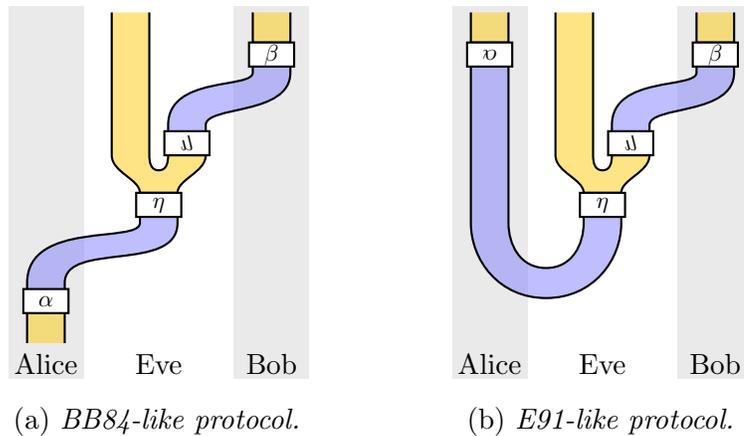

\figuretopsuck
\def \rd{0.}%
\def \scl{0.5}%
\def\sclnode{0.8}%
\def\yscl{0.8}%
\begin{calign}
\nonumber
&
\begin{tz}[scale=\scl,yscale=\yscl]
\path[background] (-1.5,-2.2) rectangle  (0.5,10.2);
\path[background] (4.5,-2.2) rectangle  (6.5,10.2);
\node at (-0.5,-1.7) {$\mathrm{Alice}$};
\node at (2.5,-1.7) {$\mathrm{Eve}$};
\node at (5.5,-1.7){$\mathrm{Bob}$};
\path[surface] (5,9+\rd) to (5,8) to [out=down, in=up, in looseness=1.5,out looseness=0.5] (2.75,6.25) to (2.75,5.25) to (3.75,5.25) to   (3.75,6.25) to [out=up, in=down, out looseness=0.6, in looseness=1.3] (6,8) to (6,9.+\rd);
\draw[edge](5,9+\rd) to (5,8) to [out=down, in=up, in looseness=1.5,out looseness=0.5] (2.75,6.25) to (2.75,5.25);
\draw[edge] (3.75,5.25) to  (3.75,6.25) to [out=up, in=down, out looseness=0.6, in looseness=1.3] (6,8) to (6,9+\rd);
\path[surface](-1,-\rd) to (-1,1) to [out=up, in=down,out looseness=1.5, in looseness=0.5] (2, 3) to(2,4) to (3,4) to (3,3)to [out=down, in=up, out looseness=1.2] (0,1) to (0,-\rd);
\draw[edge] (-1,-\rd) to (-1,1) to [out=up, in=down,out looseness=1.5, in looseness=0.5] (2, 3) to(2,4);
\draw[edge](3,4) to (3,3)to [out=down, in=up, out looseness=1.2] (0,1) to (0,-\rd);
\path[classical,draw,edge] (-1,-1) to (-1,-\rd) to (0,-\rd) to (0,-1);
\path[classical,draw,edge] (1.25,10) to [out=down, in=up] (1.25,5.25) to [out=down, in=up, in looseness=0.8] (2,4) to (3,4)to  [out=up, in=down, out looseness=0.8] (3.75,5.25) to  (2.75,5.25) to [out=down, in=right] (2.5,4.75)  to [out=left, in=down] (2.25,5.25) to [out=up, in=down] (2.25,10);
\path[classical, draw,edge] (5,10) to (5,9+\rd) to (6,9+\rd) to (6,10);
\path[fill=white,draw,edge] (-1.1,0.0) rectangle node[scale=\sclnode] {$\alpha$} +(1.2,0.8);
\path[fill=white,draw,edge] (1.9,3.2) rectangle node[scale=\sclnode,yscale=1] {$\eta$} +(1.2,0.8);
\path[fill=white,draw,edge] (2.65,5.25) rectangle node [scale=\sclnode, yscale=-1] {$\eta$} +(1.2,0.8);
\path[fill=white,draw,edge] (4.9,8.2) rectangle node[scale=\sclnode,yscale=1] {$\beta$}  +(1.2,0.8);
\path[partition] (0.5,-1.5) to (0.5,10.5);
\path[partition](4.5,-1.5) to  (4.5,10.5);
\end{tz}
&
\def \rd{0.3}
\def \rd{0.}
\def \scl{0.5}
\def\sclnode{0.8}
\def\yscl{0.8}
\begin{tz}[scale=\scl,yscale=\yscl]
\path[background] (-1.5,-2.2) rectangle  (0.5,10.2);
\path[background] (4.5,-2.2) rectangle  (6.5,10.2);
\node at (-0.5,-1.7) {$\mathrm{Alice}$};
\node at (2.5,-1.7) {$\mathrm{Eve}$};
\node at (5.5,-1.7){$\mathrm{Bob}$};
\path[surface] (5,9+\rd) to (5,8) to [out=down, in=up, in looseness=1.5,out looseness=0.5] (2.75,6.25) to (2.75,5.25) to (3.75,5.25) to   (3.75,6.25) to [out=up, in=down, out looseness=0.6, in looseness=1.3] (6,8) to (6,9.+\rd);
\draw[edge](5,9+\rd) to (5,8) to [out=down, in=up, in looseness=1.5,out looseness=0.5] (2.75,6.25) to (2.75,5.25);
\draw[edge] (3.75,5.25) to  (3.75,6.25) to [out=up, in=down, out looseness=0.6, in looseness=1.3] (6,8) to (6,9+\rd);
\path[surface]
(0,9) to (0,3) to [out=down, in=left] (1,1.5) to [out=right, in=down] (2,3)
 to (2,4) to (3,4) to (3,3) to [out=down, in=right] (1,0.5) to [out=left, in=down] (-1,3) to (-1,9);
 \draw[edge] (0,9) to (0,3) to [out=down, in=left] (1,1.5) to [out=right, in=down] (2,3) to (2,4);
 \draw[edge](3,4) to (3,3) to [out=down, in=right] (1,0.5) to [out=left, in=down] (-1,3) to (-1,9);
 \draw[classical,draw,edge] (-1,10) to (-1,9) to (0,9) to (0,10); 
\path[classical,draw,edge] (1.25,10) to [out=down, in=up] (1.25,5.25) to [out=down, in=up, in looseness=0.8] (2,4) to (3,4)to  [out=up, in=down, out looseness=0.8] (3.75,5.25) to  (2.75,5.25) to [out=down, in=right] (2.5,4.75)  to [out=left, in=down] (2.25,5.25) to [out=up, in=down] (2.25,10);
\path[classical, draw,edge] (5,10) to (5,9+\rd) to (6,9+\rd) to (6,10);
\path[fill=white,draw,edge] (-1.1,8.2) rectangle node[scale=\sclnode,rotate=180] {$\alpha$} +(1.2,0.8);
\path[fill=white,draw,edge] (1.9,3.2) rectangle node[scale=\sclnode,yscale=1] {$\eta$} +(1.2,0.8);
\path[fill=white,draw,edge] (2.65,5.25) rectangle node [scale=\sclnode, yscale=-1] {$\eta$} +(1.2,0.8);
\path[fill=white,draw,edge] (4.9,8.2) rectangle node[scale=\sclnode,yscale=1] {$\beta$}  +(1.2,0.8);
\path[partition] (0.5,-1.5) to (0.5,10.5);
\path[partition](4.5,-1.5) to  (4.5,10.5);
\end{tz}
\\*\nonumber
&
\text{(a) \emph{BB84-like protocol.}}
&
\text{(b) \emph{E91-like protocol.}}
\end{calign}

\figurecaptionsuck
\caption{Key distribution protocols.\label{fig:KD}}
\end{figure}

\subsection{Key distribution (Figures \ref{fig:qkddetailed} and \ref{fig:KD})}

\label{sec:keydistribution}
\firstparagraph{Overview.}
Quantum key distribution (QKD)~\cite{Alleaume:2007} is one of the most important protocols in quantum information. Here we describe a classical analogue which can operate on networks of groudits. The inability of the eavesdropper to read both the logical and internal state of a groubit is exploited to enable the effect. An analysis of QKD\ using a related graphical calculus has also been performed by Coecke and Perdrix~\cite{Coecke:2012b}. We focus here on BB84-style QKD~\cite{BB84}; by dagger pivotality, the E91 variant~\cite{E91} has a similar analysis (see \autoref{fig:KD}.)

\paragraph{Program.}
The basic setup of our key distribution protocol is given in \autoref{fig:KD}(a), and is similar to the BB84 QKD protocol~\cite{BB84}. Alice and Bob have an authenticated public classical channel, and a groubit channel, which are both accessible by an adversary Eve. Alice begins with a classical bit, and chooses at random to encode it into a groubit using $\alpha =\Write$ or $\alpha = \IWrite$. She sends the groubit to Bob, perhaps using a state transfer algorithm (see Section~\ref{sec:statetransfer}), but it is intercepted by Eve, who chooses to decode the message using either $\eta = \Read$ or $\eta = \IRead$; having received a classical bit she copies it, and re-encodes a groubit using $\eta^\dag$, which she sends to Bob. When Bob receives the groubit, he decodes it using $\beta = \Read$ or $\beta=\IRead$.

\paragraph{Verification.} The analysis proceeds in just the same way as for the traditional BB84 procedure. If Alice, Bob and Eve all choose the same operation ($\alpha = \eta ^\dag = \beta ^\dag$), then it is as if Alice's choice of initial bit is copied to Eve and Bob. We analyze this scenario in \autoref{fig:qkddetailed}(a), where we choose $\alpha = \eta^\dag = \beta^\dag = \Write$; using the equations of \autoref{fig:measurement}, the equation can be verified. On the other hand, if any of the 3 parties do not choose the same operation, the diagram disconnects. We analyze this in \autoref{fig:qkddetailed}(b); using the equations of \Autoref{fig:measurement, fig:redredbiunitary}, and in particular \autoref{fig:redredbiunitary}(d), this chain of equalities can also be shown, leading us to conclude that all parties receive uncorrelated random bits. Although it matches the structure of the traditional quantum analysis, in this case it is incomplete, since we do not consider all possible actions by Eve; a deeper theory quantifying information flow through groubit networks would be needed to make a stronger statement.

\paragraph{Discussion.}
This protocol may have real-world relevance, either for key distribution within an insecure data centre based on groubit networks, or on a larger scale. Our analysis here cannot be considered a full security proof; just as with genuine quantum key distribution, there are many compounding details that would affect the real security of the procedure.

\def \scl{0.5}
\def\yscl{0.8}

\section{Mathematical foundations}
\label{sec:bicategories}

\subsection{Definition of $\GAf$}
\label{fulldefinition}

\noindent
Here we give a precise definition of the 2\-category \GAf. We assume some familiarity with the theory of 2\-categories. For a good introduction see~\mbox{\cite[Chapter 7]{Borceux:1994}}.

For any category \cat C, we write $\N : \cat C \to \cat {Set}$ for the constant functor that sends all morphisms to the identity on the set of natural numbers. Also, for any two functors $S,T: \cat C \to \cat {Set}$, we write $S \times T$ for their product in the functor category.

For a groupoid \cat{G},  we say that a functor $S:\cat{G} \to \cat{FinSet}$ is \emph{free} if for any $f, f' :g\to g'$ for which there is an $x\in S(g)$ with $S(f)(x) =S(f')(x)$ it follows that $f=f'$.
We say that a functor $S:\cat H^\op \times \cat{G} \to \cat{FinSet}$ is \emph{free} if both $S(h, -):\cat{G} \to \cat{FinSet}$ and $S(-,g): \cat{H}^\op \to \cat{FinSet}$ are free for arbitrary objects $h \in \Ob(\cat H),~g\in \Ob(\cat G)$.

\ignore{ in $\cat{G}$ if, given morphisms $f:g\to g'$ and $f':g\to g'$ in $\cat{G}$ such that there exists an $h\in \Ob (\cat H)$ and $x\in S(h,g)$ with $S(h,f)(x) = S(h,f')(x)$, then it holds that $f=f'$. Freeness in $\cat H$ is defined similarly. A functor $S:\cat{H}^{op} \times \cat{G}\to \cat{FinSet}$ is \emph{free} if it is both free in $\cat{G}$ and $\cat{H}$.}

\begin{definition}
The 2\-category $\GAf$ is built from the following structures:
\begin{itemize}
\item \textbf{objects} are finite groupoids $\cat G$, $\cat H$, ...;
\item \textbf{morphisms} $S: \cat G \hto \cat H$ are free functors $S: \cat H ^\op \times \cat G \to \cat {FinSet}$;
\item \textbf{2-morphisms} $\sigma: S \hTo T$ are natural transformations \mbox{$\sigma: S \times T \Rightarrow \N$.}
\end{itemize}
Composition of morphisms and $2$-morphisms is defined below.
\end{definition}
\noindent
It is clear that this is equivalent to the restriction to skeletal groupoids, which for simplicity we use in the main body of the paper.

The morphisms of this 2\-category are also known as \textit{profunctors} or \textit{distributors}, and there is a standard way to compose them~\cite[Proposition 7.8.2]{Borceux:1994} which we define below. The 2\-morphisms can be considered as families of spans of sets\footnote{If $\cat G$ and $\cat H$ are both the terminal groupoid $\cat 1$, then morphisms $S,T:\cat 1 \hto[] \cat 1$ are finite sets and a natural transformation $\sigma: S\hTo T$ is a function $\sigma: S\times T \to \mathbb{N}$, or equivalently a (bijection class of) \emph{span of sets} $S\leftarrow \sqcup_{s\in S, t\in T} [\sigma(s,t)] \to T$, where $[\sigma(s,t)]$ is a finite set with $\sigma(s,t) \in \mathbb{N}$ elements. Analogously, a protransformation between profunctors $\sigma:S\hTo T$ can be considered as a family of spans of sets $S_{g,h}\leftarrow X_{g,h} \rightarrow T_{g,h}$, indexed by $g\in \Ob(\cat G)$ and $h \in \Ob(\cat H)$, that is compatible with the actions of $\cat G$ and $\cat H$.  } which are compatible with the groupoid actions; we will refer to them as \emph{protransformations}. 

Spelled out explicitly, a profunctor \mbox{$P:\cat{A} \hto \cat{B}$} is a family of sets $\left(P_{b,a}\right)_{b \in \ob\cat{B},a\in \ob\cat{A}}$ equipped with functions \begin{align}\label{eq:actionprofunctor}\Hom_{\cat{A}}(a,a') &\times P_{b,a} \times \Hom_{\cat{B}}(b',b)\hspace{-1.1cm}&\to P_{b',a'}\\\nonumber &(f,x,g) &\mapsto f.x.g\end{align}
for each $a,a'\in \ob \cat{A}, b,b'\in \ob\cat{B}$ such that for all $x\in P_{b,a}$ and all $f:a\to a'$, $f': a'\to a''$, $g:b'\to b$, $g':b''\to b'$, we have the following:
\[ \id_a.x.\id_b = x\hspace{1cm}(f'f).x.(gg') = f'.(f.x.g).g' \]
A profunctor is \textit{free} just when all  groups $\Aut_{\cat{A}} (a)$ and $\Aut_{\cat{B}}(b)$ act freely on $P_{b,a}$. This structure induces obvious separate left and right actions on $P_{b,a}$ which we will also make use of.

A \emph{protransformation} $\sigma:P\hTo Q$ corresponds to a family of functions $P_{b,a} \times Q_{b,a} \to[\sigma_{b,a}] \mathbb{N}$ such that the following holds, for all $p\in P_{b,a}, q\in Q_{b,a}$, $a\to[f]a'$ and $b'\to[g] b$: 
\begin{equation}\label{eq:propprotrafo}\sigma_{b',a'}\left(f.p.g, f.q.g\right) =\sigma_{b,a}\left(p,q\right)\end{equation}
From now on we will omit the subscripts of protransformations.

\paragraph{Horizontal composition.}
The composite $Q\circ P$ of profunctors $P:\cat{A} \hto\cat{B}$ and $Q:\cat{B}  \hto \cat{C}$ corresponds to the family of sets  $\left(Q\circ P\right)_{c,a}:=\quotient{\coprod_{b\in \ob\cat{B}} P_{b,a}\times Q_{c,b}}{\sim} $  with equivalence relation as follows, for any $x\in P_{b',a}$, $y \in Q_{c,b}$, $f : b\to b'$:
\vspace{-4pt}
\begin{equation}
\label{eq:equiv}
(x.f,y) \sim (x,f.y)
\end{equation}
Given profunctors $S,S':\cat{A} \hto \cat{B}$, $T,T':\cat{B} \hto\cat{C}$ and protransformations $\sigma:S\hTo S'$ and $\tau:T\hTo T'$, the horizontal composite $\tau\circ \sigma:T\circ S \hTo T'\circ S'$ corresponds to the family of functions 
\[
\left(\tau\circ\sigma\right)_{c,a}:\left(\quotient{\coprod_{b} S_{b,a} {\times} T_{c,b}}{\sim} \right)\times \left(\quotient{\coprod_{b'} S'_{b',a} {\times} T'_{c,b'}}{\sim}\right) \to \mathbb{N}
\]
defined as follows, for $a\in \Ob(\cat{A})$, $b,b'\in \Ob(\cat{B})$, $c\in \ob\cat{C}$ and $s\in S_{b,a}$, $t\in T_{c,b}$, $s'\in S'_{b',a}$, $t'\in T'_{c,b'}$:
\begin{equation} \label{eq:horcomp1}\hspace{-0.3cm}\tau\circ \sigma \left([s,t],[s',t']\right)
:=
\textstyle\sum_{f:b\to b'}
\sigma(s,s'.f) \tau(f.t,t')\end{equation}
Here $[s,t]$ and $[s',t']$ denote equivalence classes under \eqref{eq:equiv}. In the proof of \autoref{thm:bicategory} we show that this is well defined.

\paragraph{Vertical composition.} Given profunctors $P,Q,R:\cat{A} \hto \cat{B}$ and protransformations $\sigma:P\hTo Q$ and $\tau:Q\hTo R$, the {vertical composite} $\tau\sigma$ corresponds to the family of functions $\left(\tau\sigma\right)_{b,a} : P_{b,a} \times R_{b,a} \to \mathbb{N}$ defined as follows, for $a\in\Ob(\cat{A})$,  $b \in \Ob(\cat{B})$, $p\in P_{b,a}$ and $r\in R_{b,a}$:
\begin{equation}\label{eq:vertcomp}\tau\sigma(p,r) := \textstyle\sum_{q\in Q_{b,a}} \sigma(p,q)\tau(q,r)\end{equation}

\begin{theorem}[restate=theoremGAbicategory,name={}]\label{thm:GAF2cat}
\GAf is a 2\-category.
\end{theorem}
\noindent
For \autoref{thm:GAF2cat} to hold, the 1-morphisms of $\GAf$ have to be \emph{free} profunctors. Without the freeness condition, the horizontal and vertical composite of $2$-morphisms~\eqref{eq:vertcomp} and~\eqref{eq:horcomp1} would not fulfill the interchange law. Indeed, for non-free profunctors, expression~\eqref{eq:horcomp1} overcounts, and would need to be normalized by a factor $\frac{1}{|\mathrm{Stab}(s',t')|}$, where $\mathrm{Stab}(s',t')$ denotes the stabilizer of $(s',t') \in S'_{b',a}\times T'_{c,b'}$ under the action of $\Hom_{\cat{B}}(b,b')$. The resulting $2$-category would extend $\GAf$ to non-free profunctors, would still contain the $2$-category of finite groupoids, profunctors and natural transformations as a subcategory, but would not admit a dagger structure anymore. \ignore{Alternatively, allowing for protransformations to take values in $\mathbb{R}_{\geq 0}$ instead of $\mathbb{N}$, one could use the normalization factor $\frac{}{}$, which leads to a dagger $2$-category extending $\GAf$ to non-free profunctors, but whose horizontal composition would not agree with the horizontal composition of natural transformations between non-free profunctors.
}
\subsection{Dagger pivotality}

\noindent
We defined the \emph{dagger} of a protransformation in \autoref{def:spandagger}.
\ignore{
\begin{definition}
For any protransformation $\sigma: S \hTo T$, its \textit{converse} $\sigma ^\dag : T \hTo S$ is defined as follows, for $a\in\ob\cat{A},b\in\ob\cat{B}, s\in S_{b,a},t\in T_{b,a}$:
\begin{equation}
\sigma^\dagger(t,s) := \sigma(s,t)
\end{equation}
\end{definition}
}
If $\sigma$ is interpreted as a computational process, then $\sigma ^\dag$ is interpreted as its time-reversal.

\begin{definition} For any profunctor $P:\cat{A}\hto \cat{B}$, its \emph{adjoint} is the profunctor $P^*: \cat{B} \hto \cat{A}$ given by 
\begin{equation}\cat{A}^{op} \times \cat{B}\to[\mathrm{inv}\times \mathrm{inv}] \cat{A}\times \cat{B}^{op} \to[\mathrm{swap}]\cat{B}^{op} \times \cat{A} \to[P] \cat{FinSet}\end{equation}
where $\mathrm{inv}:\cat{A}^{op} \to \cat{A}$ is the canonical isomorphism mapping every morphism to its inverse.
In terms of bimodules, given a $\cat{A}{-}\cat{B}$ bimodule $\left(P_{b,a}\right)_{a,b}$, its adjoint is defined to be the $\cat{B}{-}\cat{A}$ bimodule $\left(P_{b,a}\right)_{b,a}$ with action $(f,x,g)\mapsto g^{-1}.x.f^{-1}$.
\end{definition}

\begin{definition} Given a profunctor $P:\cat{A} \hto \cat{B}$, we define its \emph{cap} to be the protransformation $\epsilon: P^*\circ P \hTo \mathbbm{1}_{\cat{A}}$ defined as follows, for $a,a'\in \Ob (\cat{A})$, $b\in \Ob(\cat{B})$, $p \in P_{b,a},p'\in P_{b,a'}, f\in \Hom(a',a)$: 
\begin{equation}\label{eq:pivotalcap}\epsilon([p,p'],f) := \delta_{f^{-1}.p,p'}
\end{equation}
We define its \emph{cup} to be the protransformation $\eta:\mathbbm{1}_{\cat{B}}\hTo P\circ P^*$ defined as follows, for $a,a'\in \ob\cat{A}, b\in \ob\cat{B}, p'\in P_{a',b}, p \in P_{a,b}$ and $f\in \Hom(a,a')$:
\begin{equation}\label{eq:pivotalcup}\eta(f,[p',p]) = \delta_{p',p.f^{-1}}
\end{equation}
\end{definition}

\begin{theorem}[restate=theoremdaggerpivotal,name={}]
\label{thm:bicategory}
The structures defined above yield a dagger pivotal structure~\cite[Section 2.1]{Carqueville:2016} for the 2\-category $\GAf$.
\end{theorem}

\subsection{Quantization}
\label{sec:quantization}

\noindent
Here we describe a \textit{quantization} procedure, in the form of a dagger $2$-functor $$\mbox{$Q : \GAf \to \cat{2Hilb}$},$$ where \cat{2Hilb} is a 2\-category that is suitable for the description of quantum computational processes~\cite{Baez:1997, Vicary:2012, Vicary:2012hq, Reutter:2016}. For this quantization procedure to be functorial, it is essential that the group actions arising from the 1\-morphisms of \GAf are free.

We take the perspective on higher vector spaces arising from algebras, bimodules and intertwiners~\cite{Baez:2012, BDSV}.
The $2$-functor
\begin{equation}
\label{eq:equivalenceAlg}
\mathrm{Rep}: \cat{C^*Alg} \to \cat{2Hilb}
\end{equation}
from the 2\-category $\cat{C^*Alg}$ of finite dimensional $C^*$\-algebras, Hilbert bimodules and intertwiners~\cite{Buss:2012} to the 2\-category $\cat{2Hilb}$ of finite dimensional 2-Hilbert spaces, linear functors and natural transformations is a dagger equivalence.

In the following, we will construct a quantization $2$-functor $Q:\GAf\to \cat{C^*Alg}.$
\begin{definition} Given a finite groupoid $\cat{G}$, its groupoid algebra (or convolution algebra) $\mathbb{C}\cat{G}$ is the vector space freely generated by all morphisms in $\cat{G}$ with algebra structure $f \star g := fg$ when $f, g$ are composable, and 0 otherwise, and $f^* := f^{-1}$.
\end{definition}
The groupoid algebra $\mathbb{C}\cat{G}$ generalizes the notion of a group algebra - for a skeletal groupoid $\cat{G}$ it is $\bigoplus_{a\in \ob \cat{G}} \Aut(a)$. In general, groupoid algebras are finite direct sums of tensor products of group algebras and matrix algebras. In particular, they are finite dimensional $C^*$\-algebras.

\begin{definition}\label{def:quantization}
The \textit{quantization} dagger $2$-functor$$Q : \GAf \to \cat{C^*Alg}$$ is defined as follows:
\begin{itemize}
\item on objects, for a groupoid \cat A, we have $Q(\cat A) := \mathbb{C}\cat A$;
\item on morphisms, for a profunctor $P: \cat{A} \hto \cat{B}$, we define $Q(P)$ to be the bimodule $\bigoplus_{b\in \ob\cat{B},a\in \ob \cat{A}} \mathbb{C}P_{b,a}$;
\item on 2-morphisms, for a protransformation $\sigma: P \hTo Q$, we define $Q(\sigma)$ as the intertwiner 
\[
\textstyle\bigoplus_{b,a}\mathbb{C}P_{b,a} \to \bigoplus_{b',a'} \mathbb{C}Q_{b',a'}\]
defined at stage $(b,a)$ as the linear map extending 
\[Q(\sigma)_{b,a} (p) = \textstyle \sum_{q\in Q_{b,a}} \sigma(p,q) q\hspace{0.5cm} p \in P_{b,a}.\]
\end{itemize}
\end{definition}

\noindent
In particular, composing this functor with the equivalence \eqref{eq:equivalenceAlg} we obtain a $2$-functor 
\[Q':\GAf \to \cat{2Hilb}\]
mapping groupoids $\cat{G}$ to their representation categories $\Rep(\mathbb{C}\cat{G}) \cong \Rep(\cat{G}) =[\cat{G},\cat{Hilb}]$.

\begin{theorem}[restate=theorempseudofunctor,name={}] \autoref{def:quantization} defines a dagger $2$-functor.
\end{theorem}

We can use these quantization results to explain the connection of our work to \textit{Hadamard matrices}\footnote{A Hadamard matrix is a unitary matrix for which every coefficient has the same absolute value; such a matrix contains the same information as a pair of mutually unbiased bases.}: under the image of the quantization functor $Q'$, every groubit yields a Hadamard matrix. To show this, we assume some knowledge of the 2\-category \cat{2Hilb}, as can be found in the references at the start of this section.

In this 2\-category, a 1\-morphism can be regarded as a matrix of Hilbert spaces, and we call a 1\-morphism \textit{nondegenerate} when all these Hilbert spaces are 1\-dimensional. It was first shown by Jones~\cite{Jones:1989}, and explored in more detail by the second author~\cite[Theorem~4.6]{Vicary:2012}, that when restricting to nondegenerate 1\-morphisms, and where the blank region labels the 1\-dimensional 2--Hilbert space, biunitaries of the form \eqref{eq:biunitary} correspond exactly to Hadamard matrices. When the groupoid \cat G is abelian (that is, a union of abelian groups), it can be checked that the boundaries \eqref{eq:boundaries} become nondegenerate 1\-morphisms under the action of the quantization functor $Q'$. Furthermore, it is clear that dagger 2\-functors preserve biunitary structures, since biunitarity is an equational property involving composition and the dagger.

With these pieces in place, the result follows: for any groubit defined over an abelian groupoid $\cat G$, its image in \cat{2Hilb} under the quantization functor $Q'$ yields a Hadamard matrix, acting on a Hilbert space given by the groupoid algebra of $\cat G$. For example, for the groubit described in Definition~\ref{def:groubit}, this yields the following Hadamard matrix:
\begin{equation}
Q'(\Swap) = \begin{pmatrix*}[r]
1 & 1 & 1 & 1
\\
1 & 1 & -1 & -1
\\
1 & -1 & 1 & -1
\\
1 & -1 & -1 & 1
\end{pmatrix*}
\end{equation}
This operates on a 4\-dimensional Hilbert space, the groupoid algebra of $\Z_2 + \Z_2$, and the matrix is written in the character basis.

\appendix

\section{Omitted proofs}

\theoremGAbicategory*
\begin{proof}
We first observe that a \emph{natural transformation} $\alpha: P \To Q$ corresponds to a family of functions $P_{b,a} \to[\alpha_{b,a}] Q_{b,a}$ such that 
\[\alpha_{b',a'} (f.p.g) = f.\alpha_{b,a} (p).g\]
for all $p \in P_{b,a}$ and $a\to[f]a',b'\to[g]b$.
Every natural transformation $\alpha$ induces a protransformation $\alpha_*(p,q) = \delta_{\alpha(p),q}$.
\paragraph{Vertical composition.}
Vertical composition \eqref{eq:vertcomp} is clearly associative with unit $1_P (x,y) =\delta_{x,y}$, and it extends the vertical composition of natural transformations.

\paragraph{Horizontal composition.}
\textit{Claim.} The composite of free profunctors is free.
\\
\textit{Proof.} The composite of free profunctors $P:\cat{A} \hto\cat{B}$ and $Q:\cat{B} \hto \cat{C}$ is given by the bimodule $\quotient{\coprod_{b\in \ob\cat{B}} P_{b,a}\times Q_{c,b}}{\sim} $ quotiented by \eqref{eq:equiv} with the obvious left and right action. Let $[p,q]$ be an equivalence class in this set and let $f\in \Aut(a)$ with $f.[p,q]=[p,q]$ or equivalently $(f.p,q) \sim (p,q)$. Thus, there is a $g\in \Aut(b)$ such that $(f.p,q) = (p.g,g^{-1}.q)$. Since $\Aut(b)$ acts freely on $Q_{c,b}$ it follows that $g=1_b$ and thus that $f.p=p.$ Since $\Aut(a)$ acts freely on $P_{b,a}$ it follows that $f=1_a$. In particular, $\Aut(a)$ acts freely on $\left(Q\circ P\right)_{c,a}$. The proof that $\Aut(c)$ acts freely is analogous.

\vspace{5pt}

\noindent
\textit{Claim.} Expression \eqref{eq:horcomp1} is well defined.
\\
\textit{Proof.} We show that \eqref{eq:horcomp1} descends to a map on the quotient. Let $\smash{\widetilde{b}\to[g] b}$ and $\smash{\widetilde{b}' \to[g'] b'}$ be arbitrary morphisms. Then 
\[\begin{split}&\tau \circ \sigma ([s.g,g^{-1}.t],[s'.g',g'^{-1}.t']) \\
&\superequals{eq:horcomp1}  \sum_{\widetilde{b}\to[\widetilde{f}]\widetilde{b}'} \sigma(s.g,s'.g'\widetilde{f})\tau(\widetilde{f}g^{-1}.t,g'^{-1}.t')\\
&\superequals{eq:propprotrafo} \sum_{\widetilde{b}\to[\widetilde{f}]\widetilde{b}'} \sigma(s,s'.g'\widetilde{f}g^{-1})\tau(g'\widetilde{f}g^{-1}.t,t')\\
&= \sum_{b\to[f]b'} \sigma(s,s'.f)\tau(f.t,t')=\tau\circ \sigma([s,t],[s',t']).\end{split}\]
Horizontal composition of protransformations extends the horizontal composition of natural transformations.
\newcommand\til[1]{\vtilde[-0.4pt]{\vphantom{t}#1}\hspace{0.1em}}
\paragraph{The interchange law.}
We show that composition of morphisms is functorial. Clearly, $1_T\circ 1_S = 1_{T\circ S}$.
\noindent
\textit{Claim.} For protransformations
\[\begin{tz}
\node (1) at (0,0) {\cat{A}};
\node (2) at (2.5,0) {\cat{B}};
\node (3) at (5,0){\cat{C}};
\draw[->] (1) to node[pos=0.5,scale=0.7] {$\mathbf{|}$} node[pos=0.5, fill=white]{\small $S'\hspace{-0.2em}$}  (2);
\draw[->]  (1) to [out=50, in=130]node[pos=0.5,scale=0.7] {$\mathbf{|}$} node [above] {\small $S$} (2);
\draw[->]  (1) to [out=-50, in=-130]node[pos=0.5,scale=0.7] {$\mathbf{|}$} node [below] {\small $S''\hspace{-0.2em}$} (2);
\node [rotate=-90] (A) at (1.25,0.35) {$\hTo$};
\node at (1.5,0.35) {\small$\sigma$};
\node [rotate=-90] (A) at (1.25,-0.35) {$\hTo$};
\node at (1.5,-0.35) {\small$\tau$};
\draw[->] (2) to node[pos=0.5,scale=0.7] {$\mathbf{|}$} node[pos=0.5,fill=white]{\small $T'\hspace{-0.2em}$}  (3);
\draw[->]  (2) to [out=50, in=130]node[pos=0.5,scale=0.7] {$\mathbf{|}$} node [above] {\small $T$} (3);
\draw[->]  (2) to [out=-50, in=-130]node[pos=0.5,scale=0.7] {$\mathbf{|}$} node [below] {\small $T''\hspace{-0.2em}$} (3);
\node [rotate=-90] (A) at (3.75,0.35) {$\hTo$};
\node at (4,0.35) {\small$\mu$};
\node [rotate=-90] (A) at (3.75,-0.35) {$\hTo$};
\node at (4,-0.35) {\small$\nu$};
\end{tz}\]
it holds that $(\nu \circ \tau)(\mu \circ \sigma)= (\nu \mu)\circ (\tau \sigma)$.\\
\newcommand\tiltil[1]{\vtilde[-0.4pt]{\vtilde[-0.4pt]{\vphantom{t}#1}}\hspace{0.1em}}
\textit{Proof.} 
For a class $[s',t'] \in (T'\circ S')_{c,a}$ we define
\begin{equation}\begin{split}\label{eq:equivclass}\langle s',t'\rangle_b &= \left\{ \left(\til{s},\til{t}\right)\in S'_{b,a} \times T'_{c,b}~|~ \left(\til{s},\til{t}\right)\sim(s',t')\right\}\\
 &=\left\{\left(\til{s},\til{t}\right) \,|\,\exists b'\to[f]b \,:\,\left(\til{s}.f, f^{-1}.\til{t}\right) = (s',t') \right\}\\
&= [s',t'] \cap (S'_{b,a} \times T'_{c,b}) \end{split}\end{equation}
By the orbit-stabilizer theorem and the fact that our profunctors act freely we can express horizontal composition \eqref{eq:horcomp1} as follows:
\begin{equation} \label{eq:horcomp2}\tau\circ \sigma \left([s,t],[s',t']\right):=\hspace{-5pt}\sum_{(\til{s},\til{t}) \in\langle s',t'\rangle_b}\hspace{-5pt} \sigma\left(s,\til{s}\right) \tau\left(t,\til{t}\right).\end{equation}
Then, for $s\in S_{ba,},t\in T_{c,b}, s''\in S''_{b',a}, t''\in T''_{c,b'}$, the following holds:
\[\begin{split}(\nu \circ \tau)&(\mu \circ \sigma)([s,t],[s'',t''])\\&\superequals{eq:vertcomp}\hspace{-1em}\sum_{\substack{\text{classes}\\[1mm] [s',t'] \in (T'\circ S')_{c,a}}}\hspace{-1.5em}(\mu \circ\sigma)([s,t],[s',t']) (\nu\circ\tau)([s',t'],[s'',t''])
\\
&\superequals{eq:horcomp2}\hspace{-1em}\sum_{\substack{\text{classes}\\[1mm] [s',t'] \in (T'\circ S')_{c,a}\\(\til{s},\til{t})\in \langle s',t'\rangle_b}}\hspace{-1.5em} \sigma(s,\til{s})\mu(t,\til{t})(\nu \circ \tau)([s',t'],[s'',t''])\\
&=\hspace{-0.8em} \sum_{\til{s}\in S_{b,a},\til{t} \in T_{c,b}}\hspace{-0.9em} \sigma(s,\til{s})\mu(t,\til{t})(\nu \circ \tau)([\til{s},\til{t}],[s'',t''])\\
&\superequals{eq:horcomp2} \hspace{-0.8em} \sum_{\til{s}\in S_{b,a},\til{t} \in T_{c,b}}\hspace{-0.9em} \sigma(s,\til{s})\mu(t,\til{t})\hspace{-1em}\sum_{(\tiltil{s},\tiltil{t})\in \langle s'',t''\rangle_b}\hspace{-1em}\tau (\til{s},\tiltil{s}) \nu(\til{t},\tiltil{t})\\
& \superequals{eq:vertcomp}  \hspace{-0.8em}\sum_{(\tiltil{s},\tiltil{t})\in \langle s'',t''\rangle_b}\hspace{-0.9em}\tau\sigma(s,\tiltil{s}) \nu\mu(t,\tiltil{t})\\
&\superequals{eq:horcomp2}\left((\nu\mu)\circ (\tau \sigma)\right)([s,t],[s'',t''])\end{split}\]
\paragraph{Associator.}
Let $P:\cat{A} \hto \cat{B},~Q:\cat{B} \hto \cat{C},~R:\cat{C} \hto \cat{D}$ be profunctors. It is well known that there are natural isomorphisms (and hence proisomorphisms) \[\alpha_{P,Q,R}: R\circ (Q\circ P) \To (R\circ Q)\circ P\] fulfilling the pentagon equation. In our notation, 
\begin{equation} \label{eq:defassociator}\alpha_{P,Q,R}([[p,q],r])= [p,[q,r]]\end{equation}
for $p\in P_{b,a}, q\in Q_{c,b}$ and $r\in R_{d,c}$. It remains to show that $\alpha$ remains natural in $P,Q,R$ after extending to $\GAf.$

\noindent 
\textit{Claim.} $\alpha_{P,Q,R}$ is natural in $P,Q,R$.\\
\textit{Proof.}
Let $\sigma:P \hTo P',\tau:Q\hTo Q'$ and $\mu: R \hTo R'$ be protransformations. 
Then, the following holds for $p\in P_{b,a},q\in Q_{c,b}, r\in R_{d,c}, p' \in P_{b',a},  q'\in Q_{c',b'}, r' \in R_{d,c'}$:
\[\begin{split}&\left((\mu\circ\tau)\circ \sigma\right)\alpha_{P,Q,R}([[p,q],r], [p',[q',r']]) \\
&\superequals{eq:defassociator}((\mu\circ \tau)\circ \sigma) ([p,[q,r]], [p',[q',r']])\\
 &\superequals{eq:horcomp1} \textstyle \sum_{b'\to[f]b} \sigma(p.f,p') (\mu\circ \tau)([q,r],f.[q',r'])\\ 
 &\superequals{eq:horcomp1} \textstyle \sum_{\substack{b'\stackrel{f}{\to}b\\c'\to[g]c}}\hspace{-0.2em}\sigma(p.f,p')\tau(q.g,f.q')\mu(r,g.r')\end{split}\]
while 
\[\begin{split}&\alpha_{P',Q',R'}\left(\mu\circ(\tau\circ \sigma)\right)([[p,q],r], [p',[q',r']]) \\
&\superequals{eq:defassociator}(\mu\circ (\tau\circ \sigma)) ([[p,q],r], [[p',q'],r'])\\ 
&\superequals{eq:horcomp1} \textstyle \sum_{c'\to[g]c} (\tau\circ\sigma)([p,q].g,[p',q']) \mu(r,g.r')\\ 
&\superequals{eq:horcomp1} \textstyle \sum_{\substack{b'\to[f] b\\ c'\to[g] c}}\sigma(p.f,p')\tau(q.g,f.q')\mu(r,g.r')
\end{split}\]

\paragraph{Unitors.}
For every groupoid $\cat{A}$, we define the identity profunctor \[\mathbbm{1}_\cat{A}:\cat{A}^\op \times \cat{A} \to \cat{Set}\hspace{0.5cm}a,a'\mapsto \Hom(a,a').\] Let $P:\cat{A} \hto \cat{B}$ be a profunctor. It is well known that there are natural isomorphisms (and thus proisomorphisms) $\lambda_P: P\circ \mathbbm{1}_{\cat{A}}\hTo P$ and $\rho_P: \mathbbm{1}_{\cat{B}}\circ P\hTo P$ fulfilling the triangle equations. In our notation 
\begin{equation}\label{eq:defunitor}\lambda_P ([f,p]) :=f.p \hspace{0.75cm}\rho_P([p,g]) =p.g \end{equation}
for $f\in \Hom(a,a'),~p\in P_{b,a},~g\in \Hom(b',b)$.

\noindent
\textit{Claim.} $\rho_P$ and $\lambda_P$ are natural in $P$.\\
\textit{Proof.} Let $\sigma: P\hTo Q$ be a protransformation. For $f\in \Hom(a,a'), p\in P_{b,a}$ and $q\in Q_{b,a'}$ we have
\[\sigma\lambda_P([f,p],q) = \sigma(f.p,q)\]
and 
\renewcommand\til[1]{\vtilde[-0.4pt]{\vphantom{g}#1}\hspace{0.1em}}
\[\begin{split}&\lambda_Q(\sigma \circ 1_{\mathbbm{1}_A})([f,p],q) \\
&\superequals{eq:vertcomp} \hspace{-1.8em}\sum_{\substack{\text{classes}\\ [g,q'] \in (Q\circ \mathbbm{1}_{\cat{A}})_{b,a'}}}\hspace{-1.5em}(\sigma \circ 1_{\mathbbm{1}_{\cat{A}}})([f,p],[g,q'])\, \lambda_Q([g,q'],q)\\
&\superequals{eq:horcomp2}\hspace{-1.8em}\sum_{\substack{\text{classes}\\ [g,q'] \in (Q\circ \mathbbm{1}_{\cat{A}})_{b,a'}\\(\til{g},\til{q}) \in \langle g,q'\rangle_a}}\hspace{-1.5em}1_{\mathbbm{1}_{\cat{A}}}(f,\til{g})\, \sigma(p,\til{q}) \,\lambda_Q([\til{g},\til{q}],q)\\
&\superequals{eq:defunitor}\hspace{-0.9em}\sum_{\substack{\til{g} \in \Hom(a,a')\\ \til{q} \in Q_{b,a}}} \delta_{f,\til{g}}\,\sigma(p,\til{g})\, \delta_{\til{g}.\til{q},q}\\
&= \sigma(p,f^{-1}.q) \superequals{eq:propprotrafo} \sigma(f.p,q) = \sigma\lambda_P([f,p],q)\end{split} \]
A similar proof shows that $\rho_P$ is natural in $P$.
\end{proof}

\theoremdaggerpivotal*
\begin{proof}
We first consider the dagger structure, then the pivotal structure.

\paragraph{Dagger.}
From the symmetry of the expressions \eqref{eq:vertcomp} and \eqref{eq:horcomp1} we conclude that $\left(\sigma\tau\right)^\dagger= \tau^\dagger\sigma^\dagger$ and $\left(\sigma\circ\tau\right)^\dagger = \sigma^\dagger \circ\tau^\dagger$. All invertible natural transformations and in particular all coherence isomorphisms are unitary.

\paragraph{Pivotality.}
We show that for every profunctor $P:\cat{A} \hto \cat{B}$, the cup \eqref{eq:pivotalcup} and cap \eqref{eq:pivotalcap} are well defined. We first show that $\epsilon$ descends to a map on the quotient. Let $p\in P_{b,a}, p'\in P_{b,a'}$, $q\in P_{b',a},q'\in P_{b',a'}$ and $f\in \Hom(a',a)$ and suppose that $(p,p') \sim (q,q')$. Then, there is a $b\to[r]b'$ s.t. $(p.r,r^{-1}\star p') = (q,q')$, where $r^{-1}\star p' =p'.r$ is the dual action. Therefore \[\epsilon([q,q'],f) = \delta_{f^{-1}.q,q'} = \delta_{f^{-1}.p.r, p'.r} = \delta_{f^{-1}.p,p'}.\] 
To show that $\epsilon$ is an intertwiner we let $p,p'$ and $f$ be as above and let $a\to[g]\widetilde{a}$ and $\widetilde{a}'\to[h] a'$. Therefore, 
\[\epsilon(g.[p,p'].h,gfh) =\delta_{h^{-1}f^{-1}g^{-1}g.p,h^{-1}.p'} =\delta_{f^{-1}.p,p'}. \]
Using unitors and associators, we can show that the left transpose of any morphism $\mu: P\hTo S$, given by the composite
\[ \left(\epsilon\circ 1_{P*}\right)\left(1_{S^*} \circ \mu \circ 1_{P*}\right) \left(1_{S*}\circ \eta\right),\]
is equal to the protransformation $\mu^*:S^*\hTo P^*$ defined by the functions 
\[S_{a,b} \times P_{a,b} \to[\mathrm{swap}]P_{a,b} \times S_{a,b} \to[\mu] \mathbb{N}.\]
For $\mu=1_P: P \hTo P$ this implies the snake equation. A similar argument shows that the right tranpose is given by the same equation. This proves that our 2\-category is a dagger pivotal 2\-category.
\end{proof}

\theorempseudofunctor*

\begin{proof}[Proof (sketch)]
Here, we use the description of $\cat{C^*Alg}$ in terms of symmetric separable dagger Frobenius algebras, Frobenius bimodules and intertwiners~\cite{Vicary:2010,Heunen:2014,Carqueville:2016}.
We first show that the data given in \autoref{def:quantization} is well defined:
Given a profunctor $P:\cat{A} \hto \cat{B}$, the vector space $\bigoplus_{b,a} \mathbb{C}P_{b,a}$ is a $\mathbb{C}\cat{A}{-}\mathbb{C}\cat{B}$ bimodule with action 
\[\mathbb{C}\cat{A} \otimes \bigoplus_{b,a} \mathbb{C}P_{b,a} \otimes \mathbb{C} \cat{B} \to \bigoplus_{b,a} \mathbb{C}P_{b,a}\]
induced from the action \eqref{eq:actionprofunctor} of the morphisms on $P_{b,a}$ (again defined to be $0$ on non-compatible components).

Given a protransformation $\sigma:P\hTo T$, the linear map $Q(\sigma)$ is indeed an intertwiner: For $p\in P_{b,a}, a\to[f] a', b'\to[g]b$:
\[\begin{split}Q(\sigma) (f.p.g) &=\sum_{t\in T_{b,a}} \sigma(f.p.g,t) t = \sum_{t\in T_{b,a}} \sigma(p,f^{-1}.t.g^{-1}) t\\&= \sum_{\widetilde{t}\in T_{b,a}} \sigma(p,\widetilde{t})f.\widetilde{t}.g= f.Q(\sigma).g\end{split}\]
By definition, $Q$ preserves the dagger structure.

The critical step in the proof of functoriality is showing that the horizontal composition of 2\-morphisms is preserved.
Formally, we defined horizontal composition of profunctors $P:\cat{A} \hto \cat{B}$ and $S:\cat{B}\hto \cat{C}$ and protransformations via the coequalizer of the maps 
\begin{align}\nonumber\textstyle\coprod_{b,b'} P_{b',a} \times \Hom(b,b') \times S_{c,b} &\to \textstyle\coprod_{b} P_{b,a} \times S_{c,b} \\\nonumber(p,f,s) &\mapsto (p.f,s)\\\nonumber(p,f,s) &\mapsto (p,f.s)\end{align}
leading to the equivalence relation $(p,s) \sim (p.f,f^{-1}.s)$.
We denoted this coequalizer as follows:
\[ \epsilon: \coprod_{b\in \ob\cat{B}}P_{b,a} \times S_{c,b}\to \left(S\circ P\right)_{c,a}\hspace{1cm} (p,s) \mapsto [p,s]\]
 In \cat{C^*Alg}, the horizontal composition of $Q(P)$ and $Q(S)$ are defined via the splitting of the dagger idempotent~\cite{Heunen:2014} $X\in \End\left(\bigoplus_{b,a} \mathbb{C}P_{b,a} \otimes \bigoplus_{c,\widetilde{b}}\mathbb{C}S_{c,\widetilde{b}}\right)$ given at stage $((b,a,c,\widetilde{b}), (b',a',c',\widetilde{b}'))$ as 
  \begin{align}\nonumber \mathbb{C}P_{b,a}\otimes \mathbb{C}S_{c,\widetilde{b}} &\to \mathbb{C}P_{b',a'} \otimes \mathbb{C}S_{c',\widetilde{b}'}\\\nonumber 
  (p,s) &\mapsto \frac{\delta_{b,\widetilde{b}}\delta_{b',\widetilde{b}'}\delta_{a,a'}\delta_{c,c'}}{|\Hom(b',b)|}\sum_{b'\to[f]b} (p.f,f^{-1}.s)
  \end{align}
To show that our functor preserves horizontal composition we thus have to show that the quantization of the coequalizer $Q(\epsilon): (p,s) \mapsto [p,s]$ splits $X$ (up to a constant). And indeed, \[Q(\epsilon)^\dagger Q(\epsilon)\in \End\left(\bigoplus_{a,b,c} \mathbb{C}P_{b,a} \otimes \mathbb{C}S_{c,b}\right)\] is given at stage $((a,b,c),(a',b',c')$ as 
\begin{align}\nonumber \mathbb{C}P_{b,a} \otimes \mathbb{C}S_{c,b}&\to \mathbb{C}P_{b',a'} \otimes \mathbb{C} S_{c',b'}\\\nonumber (p,s) &\mapsto \delta_{a,a'}\delta_{c,c'}\sum_{(\widetilde{p},\widetilde{s}) \in\langle p,s\rangle_{b'}}(\widetilde{p},\widetilde{s})\\ \nonumber &= \delta_{a,a'}\delta_{c,c'} \frac{1}{|\Stab(p,s)|}\sum_{b'\to[f]b} (p.f,f^{-1}.s)\end{align}
where $\Stab(p,s)= \left\{b\to[g] b~|~ (p.g,g^{-1}.s) = (p,s)\right\}$ is the stabilizer of the joint action of $\Aut_{\cat{B}}(b)$ on $P_{b,a}\times S_{c,b}$ and in $\langle p,s\rangle_{b'}$ is defined in \eqref{eq:equivclass}. 
Since all groupoids in \GAf act freely, this stabilizer vanishes and $Q(\epsilon)^\dagger Q(\epsilon)$ is indeed proportional to $X$. Similarly $Q(\epsilon)Q(\epsilon)^\dagger$ is proportional to $\mathbbm{1}_{Q(S\circ P)}$. \end{proof}

\bibliographystyle{alpha}
\bibliography{references_LMCS}

\end{document}

%% file: header_LMCS.tex
\usepackage{amsmath,amsthm,amssymb}
\usepackage{graphicx}    
\usepackage{calc}
\usepackage{docmute}
\usepackage{array}
\usepackage{xspace}
\usepackage[numbers,sort,compress]{natbib}
\usepackage[font={it}]{caption}
\usepackage{bbm}            
\usepackage{etoolbox}      
\usepackage{hyperref}       
\usepackage{xstring} 
\usepackage{mathtools}
\usepackage{thm-restate}
\usepackage{docmute}
\def\subparagraph{}
\usepackage{enumitem}

\newtheorem{theorem}{Theorem}[section]
\newtheorem{proposition}[theorem]{Proposition}

\theoremstyle{definition}
\newtheorem{definition}[theorem]{Definition}
\newtheorem{varremark}[theorem]{Remark}

\makeatletter
\renewcommand\Autoref[1]{\@first@ref#1,@}
\def\@throw@dot#1.#2@{#1}
\def\@set@refname#1{
    \edef\@tmp{\getrefbykeydefault{#1}{anchor}{}}%
    \def\@refname{\@nameuse{\expandafter\@throw@dot\@tmp.@autorefname}s}%
}
\def\@first@ref#1,#2{%
  \ifx#2@\autoref{#1}\let\@nextref\@gobble
  \else%
    \@set@refname{#1}
    \@refname~\ref{#1}
    \let\@nextref\@next@ref
  \fi%
  \@nextref#2%
}
\def\@next@ref#1,#2{%
   \ifx#2@ and~\ref{#1}\let\@nextref\@gobble
   \else, \ref{#1}
   \fi%
   \@nextref#2%
}
\makeatother

\usepackage{tikz}

\usetikzlibrary{calc, decorations.markings, decorations.pathmorphing, intersections, shapes, fit, shapes, decorations.pathreplacing}
\pgfdeclarelayer{back}
\pgfdeclarelayer{front}
\pgfsetlayers{back,main,front}
\makeatletter
\pgfkeys{%
  /tikz/on layer/.code={\pgfonlayer{#1}\begingroup\aftergroup\endpgfonlayer\aftergroup\endgroup},
  /tikz/node on layer/.code={\gdef\node@@on@layer{\setbox\tikz@tempbox=\hbox\bgroup\pgfonlayer{#1}\unhbox\tikz@tempbox\endpgfonlayer\egroup}\aftergroup\node@on@layer},
  /tikz/end node on layer/.code={\endpgfonlayer\endgroup\endgroup}
}
\def\node@on@layer{\aftergroup\node@@on@layer}

\newenvironment{tz}
{\begin{aligned}\begin{tikzpicture}}
{\end{tikzpicture}\end{aligned}\,}
\tikzset{blob/.style={draw, circle, fill=white, inner sep=1pt, minimum width=5pt, font=\scriptsize, node on layer=front, thick}}
\tikzset{greenregion/.style={fill=green, fill opacity=0.3, draw=none}}
\tikzset{redregion/.style={fill=red, fill opacity=0.3, draw=none}}
\tikzset{blueregion/.style={fill=blue, fill opacity=0.3, draw=none}}
\tikzset{yellowregion/.style={fill=yellow, fill opacity=0.5, draw=none}}
\tikzset{cyanregion/.style={fill=cyan, fill opacity=0.3, draw=none}}
\tikzset{orangeregion/.style={fill=orange, fill opacity=0.6, draw=none}}
\tikzset{solidgreenregion/.style={fill=green!30, fill opacity=1, draw=none}}
\tikzset{solidredregion/.style={fill=red!30, fill opacity=1, draw=none}}
\tikzset{solidblueregion/.style={fill=blue!30, fill opacity=1, draw=none}}
\tikzset{solidyellowregion/.style={fill=yellow!30, fill opacity=1, draw=none}}
\tikzset{string/.style={line width=0.7pt}}
\tikzset{zig/.style={decoration={zigzag,segment length=3, amplitude=0.5}}}
\tikzset{bnd/.style={draw,string}}   
\tikzset{projector/.style={circle, draw, font=\scriptsize, inner sep=-5pt, minimum width=0.35cm, string, fill=white}}
\tikzset{dimension/.style={font=\scriptsize, inner sep=1pt}}
\tikzset{edge/.style={draw,thick}}
\tikzset{surface/.style={draw=none, fill=blue!50, fill opacity=0.5}}
\tikzset{classical/.style={draw=none, fill=red!50, fill opacity=0.5}}
\tikzset{background/.style={fill=gray!40, fill opacity=0.4}}
\tikzset{partition/.style={}}
\tikzset{hor/.style={draw, dashed, ultra thin ,opacity=1}}

\makeatletter
\def\calign@preamble{&\hfil\strut@\setboxz@h{\@lign$\m@th\displaystyle{##}$}\ifmeasuring@\savefieldlength@\fi\set@field\hfil\tabskip\alignsep@}
\let\cmeasure@\measure@
\patchcmd\cmeasure@{\divide\@tempcntb\tw@}{}{}{}
\patchcmd\cmeasure@{\divide\@tempcntb\tw@}{}{}{}
\patchcmd\cmeasure@{\ifodd\maxfields@\global\advance\maxfields@\@ne\fi}{}{}{}
\newenvironment{calign}{\let\align@preamble\calign@preamble\let\measure@\cmeasure@\align}{\endalign}
\makeatother

\makeatletter
\let\oldmarginpar\marginpar
\renewcommand*{\marginpar}[1]{%
   \begingroup%
     \if@firstcolumn\else\reversemarginpar\fi
   \oldmarginpar{#1}%
   \endgroup%
}
\makeatother
\newcounter{jvcommcounter}
\newcounter{drcommcounter}
\newcommand{\End}{\mathrm{End}}
\newcommand{\Hom}{\mathrm{Hom}}
\newcommand{\Aut}{\mathrm{Aut}}
\newcommand\cat[1]{\ensuremath{\mathbf{#1}}}

\newcommand\G{\ensuremath{\cat{G}}\xspace}
\newcommand\B{\ensuremath{\mathcal{B}}\xspace}
\newcommand\Z{\ensuremath{\mathbb{Z}}\xspace}

\newcommand\superequals[1]{\stackrel {\makebox[0pt]{\tiny\eqref{#1}}} =}

\newcommand{\quotient}[2]{\left.\raisebox{.1em}{$#1$}\middle/\raisebox{-.1em}{$#2$}\right.}
\newcommand\eqgap{\hspace{1pt}}
\newcommand\N{\mathbb{N}}
\def\hrt{0.707107} 

\newcommand\vc[1]{\begin{tabular}{ccccccc}#1\end{tabular}}

\newcommand\ignore[1]{}

\def\maskpointheight{8pt}
\tikzset{mask point/.style={transform shape, sloped, 
minimum width=15pt, minimum height=\maskpointheight, inner sep=0pt, ultra thin, font=\tiny}}

\tikzset{
clip even odd rule/.code={\pgfseteorule},
invclip/.style={clip,insert path=[clip even odd rule]{
   [reset cm](-\maxdimen,-\maxdimen)rectangle(\maxdimen,\maxdimen)
    }}}

\usepackage[justification=centering]{caption}

\renewcommand\paragraph[1]{

\vspace{5pt}\noindent\textbf{#1}}

\newcommand\firstparagraph[1]{
\noindent
\textbf{#1}}

\def\figuretopsuck{\vspace{0pt}} 
\def\figurecaptionsuck{\vspace{0pt}} 
\def\figurecaptionpostsuck{\vspace{0pt}} 

\newcommand\Read{\textbf{Read}\xspace}
\newcommand\Write{\textbf{Write}\xspace}
\newcommand\Swap{\textbf{Swap}\xspace}
\newcommand\Init{\textbf{Init}\xspace}
\newcommand\Tick{\textbf{Tick}\xspace}
\newcommand\Erase{\textbf{Erase}\xspace}
\newcommand\Split{\textbf{Split}\xspace}
\newcommand\Rand{\textbf{Rand}\xspace}
\newcommand\CTick{\textbf{CTick}\xspace}
\newcommand\IRead{\textbf{IRead}\xspace}
\newcommand\IWrite{\textbf{IWrite}\xspace}
\newcommand\LRead{\textbf{LRead}\xspace}
\newcommand\RRead{\textbf{RRead}\xspace}
\newcommand\F{\ensuremath{F}}
\newcommand\obG{\ensuremath{\mathrm{Ob}(\G)}}
\newcommand\morG{\ensuremath{\mathrm{Mor}(\G)}}

\makeatletter
\def\slashedarrowfill@#1#2#3#4#5{%
  $\m@th\thickmuskip0mu\medmuskip\thickmuskip\thinmuskip\thickmuskip
\relax#5#1\mkern-7mu%
\cleaders\hbox{$#5\mkern-2mu#2\mkern-2mu$}\hfill
\mathclap{#3}\mathclap{#2}%
\cleaders\hbox{$#5\mkern-2mu#2\mkern-2mu$}\hfill
\mkern-7mu#4$%
}
\def\rightslashedarrowfill@{%
  \slashedarrowfill@\relbar\relbar\mapstochar\rightarrow}
\newcommand\xslashedrightarrow[2][]{%
  \ext@arrow 0055{\rightslashedarrowfill@}{#1}{#2}}
\def\Rightslashedarrowfill@{%
  \slashedarrowfill@\relbar\relbar\mapstochar\Rightarrow}
\newcommand\xslashedRightarrow[2][]{%
  \ext@arrow 0055{\Rightslashedarrowfill@}{#1}{#2}}
\makeatother

\renewcommand{\to}[1][]{\ensuremath{\xrightarrow{#1}}}
\newcommand{\hto}[1][]{\ensuremath{\xslashedrightarrow{#1}}}
\newcommand{\hTo}[1][]{\ensuremath{\xslashedRightarrow{#1}}}
\newcommand{\To}[1][]{\ensuremath{\xRightarrow{#1}}}

\newcommand\op{\ensuremath{\mathrm{op}}}
\newcommand\Ob{\ensuremath{\mathrm{Ob}}}
\newcommand\Mor{\ensuremath{\mathrm{Mor}}}

\newcommand\GAf{\ensuremath{\cat{GpdAct_{f}}}\xspace}

\newcommand{\Stab}{\ensuremath{\mathrm{St}}}

\newcommand\Rep{\ensuremath{\mathrm{Rep}}}

\newcommand{\ob}{\ensuremath{\mathrm{ob}}}
\newcommand{\id}{\ensuremath{\mathrm{id}}}
\newcommand*{\vtilde}[2][0pt]{
  \setbox0=\hbox{$#2$}%
  \widetilde{\mathrlap{\phantom{\rule{\wd0}{\ht0+{#1}}}}\smash{#2}}%
}

\def\hrt{0.707107}

\def\soplus{{\oplus}}
\tikzset{programlabel/.style={anchor=west, align=left, scale=0.75}}
\tikzset{protocoldiagram/.style={yscale=0.7, xscale=0.7}}
\allowdisplaybreaks


%% file: paper4_LMCS.bbl
\begin{thebibliography}{BDSPV15}

\bibitem[AM15]{Adesanmi_2015}
Akintomide Adesanmi and Lotfi Mhamdi.
\newblock Controlling {TCP} {I}ncast congestion in data centre networks.
\newblock In {\em {ICCW} 2015}. {IEEE}, 2015.

\bibitem[Bae97]{Baez:1997}
John~C. Baez.
\newblock Higher-dimensional algebra {II}. 2-{Hilbert} spaces.
\newblock {\em Advances in Mathematics}, 127(2):125--189, 1997.
\newblock \href{https://arxiv.org/9609018}{\nolinkurl{arXiv:9609018}}.

\bibitem[BB85]{BB84}
Charles~H. Bennett and Gilles Brassard.
\newblock Quantum public key distribution.
\newblock {\em IBM Tech. Disc. Bul.}, 28:3153--3163, 1985.

\bibitem[BBFW12]{Baez:2012}
John Baez, Aristide Baratin, Laurent Freidel, and Derek Wise.
\newblock Infinite-dimensional representations of 2-groups.
\newblock {\em Memoirs of the American Mathematical Society}, 219(1032), 2012.

\bibitem[BD01]{Baez:2001}
John Baez and James Dolan.
\newblock From finite sets to {F}eynman diagrams.
\newblock In {\em Mathematics {Unlimited}}, pages 29--50. 2001.

\bibitem[BD15]{Backens:2015}
Miriam Backens and Ali~Nabi Duman.
\newblock A complete graphical calculus for {S}pekkens' toy bit theory.
\newblock {\em Foundations of Physics}, 46(1):70--103, 2015.

\bibitem[BDSPV15]{BDSV}
Bruce Bartlett, Christopher~L. Douglas, Christopher Schommer-Pries, and Jamie
  Vicary.
\newblock Modular categories as representations of the 3-dimensional bordism
  2-category.
\newblock 2015.
\newblock \href{http://arxiv.org/abs/1509.06811}{arXiv:1509.06811}.

\bibitem[Ber]{Bernstein:2009}
Daniel~J. Bernstein.
\newblock Introduction to post-quantum cryptography.
\newblock In {\em Post-Quantum Cryptography}, pages 1--14. Springer Nature.

\bibitem[BHW10]{Baez:2010}
John Baez, Alexander Hoffnung, and Christopher Walker.
\newblock {HDA VII}: {G}roupoidification.
\newblock {\em {TAC}}, 24(18):489--553, 2010.

\bibitem[BLLR97]{Bergeron_1997}
Francois Bergeron, Gilbert Labelle, Pierre Leroux, and Margaret Readdy.
\newblock {\em Combinatorial Species and Tree-like Structures}.
\newblock Cambridge University Press ({CUP}), 1997.

\bibitem[BMZ12]{Buss:2012}
Alcides Buss, Ralf Meyer, and Chenchang Zhu.
\newblock A higher category approach to twisted actions on {C}{${}^*$}
  -algebras.
\newblock {\em Proceedings of the Edinburgh Mathematical Society},
  56(02):387--426, Aug 2012.

\bibitem[Bor94]{Borceux:1994}
Francis Borceux.
\newblock {\em Handbook of Categorical Algebra}.
\newblock CUP, 1994.

\bibitem[Bor16]{Borrill:2016}
Paul Borrill.
\newblock The timeless datacentre.
\newblock {\em Stanford Colloquium on Computer Systems}, 2016.
\newblock
  \href{https://www.youtube.com/watch?v=IPTlTmH-YvQ}{YouTube:IPTlTmH-YvQ}.

\bibitem[BV14]{Bar:2014}
Krzysztof Bar and Jamie Vicary.
\newblock Groupoid semantics for thermal computing.
\newblock \href{https://arxiv.org/abs/1401.3280}{arXiv:1401.3280}, 2014.

\bibitem[CES11]{Coecke_2011}
Bob Coecke, Bill Edwards, and Robert~W. Spekkens.
\newblock Phase groups and the origin of non-locality for qubits.
\newblock {\em ENTCS}, 270(2):15--36, 2011.

\bibitem[CP12]{Coecke:2012b}
Bob Coecke and Simon Perdrix.
\newblock Environment and classical channels in categorical quantum mechanics.
\newblock {\em LMCS}, 8(4), 2012.

\bibitem[CR16]{Carqueville:2016}
Nils Carqueville and Ingo Runkel.
\newblock Orbifold completion of defect bicategories.
\newblock {\em Quantum Topol.}, 7(2):203--279, 2016.
\newblock \href{https://arxiv.org/1210.6363}{\nolinkurl{arXiv:1210.6363}}.

\bibitem[DM16]{Disilvestro:2016}
Leonardo Disilvestro and Damian Markham.
\newblock Quantum protocols within {S}pekkens' toy model.
\newblock \href{https://arxiv.org/abs/1608.09012}{arXiv:1608.09012}, 2016.

\bibitem[ea14]{Alleaume:2007}
Romain~All\'eaume et~al.
\newblock Using quantum key distribution for cryptographic purposes: a survey.
\newblock {\em {TCS}}, 560(1):62--81, 2014.

\bibitem[Eke91]{E91}
Artur Ekert.
\newblock Quantum cryptography based on {B}ell's theorem.
\newblock {\em Physical Review Letters}, 67:661, 1991.

\bibitem[HVW14]{Heunen:2014}
Chris Heunen, Jamie Vicary, and Linde Wester.
\newblock Mixed quantum states in higher categories.
\newblock {\em Electronic Proceedings in Theoretical Computer Science},
  172:304--315, Dec 2014.

\bibitem[IAC99]{Iren_1999}
Sami Iren, Paul~D. Amer, and Phillip~T. Conrad.
\newblock The transport layer: tutorial and survey.
\newblock {\em {ACM} Computing Surveys}, 31(4):360--404, 1999.

\bibitem[JLW16]{Jaffe:2016a}
Arthur Jaffe, Zhengwei Liu, and Alex Wozniakowski.
\newblock Holographic software for quantum networks.
\newblock 2016.
\newblock \href{https://arxiv.org/abs/1605.00127}{arXiv:1605.00127}.

\bibitem[JMS13]{Jones:2013}
Vaughan F.~R. Jones, Scott Morrison, and Noah Snyder.
\newblock The classification of subfactors of index at most 5.
\newblock {\em Bull. Amer. Math. Soc.}, 51(2):277--327, 2013.
\newblock \href{https://arxiv.org/1304.6141}{\nolinkurl{arXiv:1304.6141}}.

\bibitem[Jon89]{Jones:1989}
Vaughan~F.R. Jones.
\newblock On knot invariants related to some statistical mechanical models.
\newblock {\em Pac. J. Math.}, 137:311--334, 1989.

\bibitem[Jon99]{Jones:1999}
Vaughan F.~R. Jones.
\newblock {Planar algebras, I}.
\newblock 1999.
\newblock \href{https://arxiv.org/abs/math/9909027}{arXiv:math/9909027}.

\bibitem[Joy81]{Joyal_1981}
Andr{\'{e}} Joyal.
\newblock Une th{\'{e}}orie combinatoire des s{\'{e}}ries formelles.
\newblock {\em Advances in Mathematics}, 42(1):1--82, 1981.

\bibitem[Mau93]{Maurer:1993}
Ueli~M. Maurer.
\newblock Protocols for secret key agreement by public discussion based on
  common information.
\newblock {\em IEEE Transactions on Information Theory}, 39(3):733--742, 1993.

\bibitem[Mor06]{Morton:2006}
{\em Theory Appl. Categ.}, 16:No. 29, 785--854, 2006.
\newblock \href{http://arxiv.org/abs/math/0601458}{arXiv:math/0601458}.

\bibitem[MP14]{Morrison:2014}
Scott Morrison and Emily Peters.
\newblock The little desert? {S}ome subfactors with index in the interval $(5,
  3 + \sqrt 5)$.
\newblock {\em International Journal of Mathematics}, 25(08):1450080, 2014.
\newblock \href{https://arxiv.org/abs/1205.2742}{arXiv:1205.2742}.

\bibitem[NC09]{Nielsen:2009}
Michael Nielsen and Isaac Chuang.
\newblock {\em Quantum Computation and Quantum Information}.
\newblock CUP, 2009.

\bibitem[Ocn89]{Ocneanu:1989}
Adrian Ocneanu.
\newblock Quantized groups, string algebras, and {Galois} theory for algebras.
\newblock In {\em Operator Algebras and Applications}, pages 119--172. {CUP},
  1989.

\bibitem[Pus12]{Pusey:2012}
Matthew~F. Pusey.
\newblock Stabilizer notation for spekkens' toy theory.
\newblock {\em Foundations of Physics}, 42(5):688--708, 2012.

\bibitem[RV16]{Reutter:2016}
David Reutter and Jamie Vicary.
\newblock {Biunitary constructions in quantum information}.
\newblock 2016.
\newblock \href{https://arxiv.org/abs/1609.07775}{arXiv:1609.07775}.

\bibitem[RV18]{Reutter:2017a}
David Reutter and Jamie Vicary.
\newblock Shaded tangles for the design and verification of quantum programs.
\newblock 2018.
\newblock \href{https://arxiv.org/abs/1805.01540}{arXiv:1805.01540}.

\bibitem[Sel10]{Selinger:2010}
Peter Selinger.
\newblock A survey of graphical languages for monoidal categories.
\newblock In {\em N. Struct. Phys.}, pages 289--355. Springer, 2010.
\newblock \href{https://arxiv.org/abs/0908.3347}{arXiv:0908.3347}.

\bibitem[Spe07]{Spekkens:2007}
Robert~W. Spekkens.
\newblock Evidence for the epistemic view of quantum states: A toy theory.
\newblock {\em PRA}, 75(3), 2007.

\bibitem[Vic10]{Vicary:2010}
Jamie Vicary.
\newblock Categorical formulation of finite-dimensional quantum algebras.
\newblock {\em Communications in Mathematical Physics}, 304(3):765--796, Nov
  2010.

\bibitem[Vic12a]{Vicary:2012hq}
Jamie Vicary.
\newblock Higher quantum theory.
\newblock 2012.
\newblock \href{https://arxiv.org/abs/1207.4563}{arXiv:1207.4563}.

\bibitem[Vic12b]{Vicary:2012}
Jamie Vicary.
\newblock Higher semantics of quantum protocols.
\newblock In {\em Proceedings of LICS 2012}, 2012.

\bibitem[VV14]{Vazirani_2014}
Umesh Vazirani and Thomas Vidick.
\newblock Robust device independent quantum key distribution.
\newblock In {\em ITCS 2014}, 2014.

\bibitem[Zha15]{Zhao:2015}
Yongwang Zhao.
\newblock A survey on formal specification and verification of separation
  kernels.
\newblock 2015.

\end{thebibliography}
